\documentclass[conference]{IEEEtran}
\IEEEoverridecommandlockouts
\usepackage{cite}
\usepackage{amsmath,amssymb,amsfonts}
\usepackage{algorithmic}
\usepackage{graphicx}
\usepackage{textcomp}
\usepackage{xcolor}
\usepackage{comment}
\usepackage{subcaption}
\usepackage{pgfplotstable}
\usepackage{booktabs}

\usepackage{array}
\newcolumntype{C}{>{\raggedleft\arraybackslash}m{1.5cm}}
\newcolumntype{P}{>{\raggedright\arraybackslash}p{3.5cm}}

\def\BibTeX{{\rm B\kern-.05em{\sc i\kern-.025em b}\kern-.08em
    T\kern-.1667em\lower.7ex\hbox{E}\kern-.125emX}}

\newcommand{\aac}[1]{{\textcolor{blue}{AAC: #1}}}
\newcommand{\excise}[1]{{}}

\begin{document}


\title{Modeling the Carbon Footprint of HPC: The Top 500 and  EasyC\\
}

\author{\IEEEauthorblockN{1\textsuperscript{st} Varsha Rao}
\IEEEauthorblockA{\textit{University of Chicago}\\
Chicago, IL, USA \\
varsharao@uchicago.edu}
\and
\IEEEauthorblockN{2\textsuperscript{nd} Andrew A. Chien}
\IEEEauthorblockA{
\textit{University of Chicago \& Argonne National Laboratory}\\
Chicago, IL, USA \\
aachien@uchicago.edu}
}

\maketitle

\begin{abstract}
Climate change is a critical concern for HPC systems, but GHG protocol carbon-emission accounting methodologies are difficult for a single system, and effectively infeasible for a collection of systems.
As a result, there is no HPC-wide carbon reporting, and even the largest HPC sites do not do GHG protocol reporting.


We assess the carbon footprint of HPC, focusing on the Top 500 systems.\footnote{This work has been accepted at the ACM/IEEE Workshops of the International Conference for High Performance Computing, Networking, Storage and Analysis (SC Workshops '25), November 16--21, 2025, St Louis, MO, USA.}  The key challenge lies in 
modeling the carbon footprint with limited data availability. 

With the disclosed Top500.org data, and using a new tool, EasyC, we were able to model the operational carbon of 391 HPC systems and the embodied carbon of 283 HPC systems. 
We further show how this coverage can be enhanced by exploiting additional public information.  With improved coverage, then interpolation is used to produce the first carbon footprint estimates of the Top 500 HPC systems.  They are 1.4 million MT CO2e operational carbon (1 Year) and 1.9 million MT CO2e embodied carbon.  We also project how the Top 500's carbon footprint will increase through 2030.

A key enabler is the EasyC tool which models  carbon footprint with only a few data metrics.  We explore availability of data and enhancement, showing that coverage can be increased to 98\% of Top 500 systems for operational and 80.8\% of the systems for embodied emissions. 

\end{abstract}


\section{Introduction}
\label{intro}
Global warming is accelerating at an alarming rate\cite{UN_climate_change_2025}, and computing's increased use is an important and growing contributor. Data center global power demand is predicted to increase 165\% by 2030 compared to 2023\cite{ai_goldman_sachs}.
A significant part of this growth is driven by emerging compute and hardware-intensive applications, such as large language models \cite{ai_exploding_topics,ai_forbes, ai_australian_tech} and the increasing use of scientific simulations\cite{hpc_outlook}. While these applications offer substantial potential for scientific progress, they also carry a considerable carbon footprint.
Within the High Performance Computing (HPC) community, there is rising concern as evidenced by increased scientific meetings, and a growing body of literature \cite{sc_sustainability,pasc_sustainability,sop_sustainability,li2023sustainablehpc,chadha_sustainability_hpc,benhari2025top500}.  This interest is driven both by the civic/environmental awareness of the HPC community, and more starkly by the rapid growth of the carbon footprint of HPC systems which are consuming more energy and producing higher carbon emissions due to continued expansion \cite{top500_2025,top500_2024}. This concern is broad-based, despite little broad analysis of HPC carbon footprint.





Despite this, carbon emission reporting for HPC systems is rare.  For example, none of the Top 25 computing systems (Top 500 list) report their carbon footprint.  GHG protocol carbon emission accounting methodologies \cite{ghg_protocol2024} require far too much effort, and more problematically a broad collection of accurate data inputs.
Despite that, GHG protocol approaches are used by many large companies, including tech companies\cite{amazon_sustainability_2024,google_sustainability_2024,microsoft_sustainability_2024,meta_sustainability_2024}, at large cost and with outcomes of questionable value.

Academic research studies have skirted the limits of significant per-system effort and extensive data requirements (under the GHG protocols), compromising either the rigor of studies, guessing or estimating many of the input data metrics  \cite{chadha_sustainability_hpc,li2023sustainablehpc,gupta2022act}. External estimates of computer systems data metrics are even more difficult, making objective academic study challenging and often inaccurate.  So, these existing examples provide no way to report the carbon footprint of a specific identifiable set of systems, such as the Top500, government-funded research computing facilities, or scientific HPC platforms in Europe.
%
%
This is unfortunate because to reduce emissions from HPC systems, we need widespread carbon footprint assessments. Achieving this at scale requires methods that work with limited data and low human effort.



We use the Top500 November 2024 list\cite{top500_2024} as a proxy to evaluate the carbon emissions of HPC systems.  We perform a broad analysis of the Top 500 systems, enabled by the  EasyC\cite{easyc_pearc} tool that dramatically reduces the detailed data requirements and effort, compared to the GHG protocol.
The results characterize the operational and embodied carbon for this collection of systems, representative of the HPC computing community.
We further examine multiple data input scenarios, assessing coverage and the change that new data can make to carbon footprint models (sensitivity).  Finally, we use the data to project the future changes in HPC carbon footprint, and important metrics such as compute-to-carbon footprint ratio.


Specific contributions include:

\begin{itemize}
\item The first assessment of Top500 carbon footprint. Annual operational carbon is 1.39 million MT CO2e (equivalent to 325k gasoline-powered vehicles annual emissions). Embodied carbon is 1.88 million MT CO2e (equivalent to 439k gasoline-powered vehicles annual emissions).

\item An assessment of a carbon modeling tool, EasyC, that shows the carbon footprint (operational and embodied) of 56.6\% of the Top 500 systems can be captured using only the data from Top500.org.

\item With additional public information, the coverage increases to 98\% for operational carbon and 80.8\% for embodied carbon. The carbon footprint of the remaining systems can be further computed through interpolation. Thus the carbon footprint of the entire Top500 can be computed with Top500.org data, other publicly available sources, and interpolation.


\item Projection of the Top 500 carbon footprint thru 2030.  Operational carbon footprint continues to increase rapidly (10.3\%/year), reaching 1.8x its 2024 level by 2030.  Embodied carbon grows slower (2\%/year), reaching 1.1x.
\end{itemize}

\section{Problem}
\label{problem}





Carbon emissions from HPC systems are increasing, but sustainability reporting for these systems is basically non-existent.  Existing carbon accounting methodologies require substantial data and human effort, making them impractical for widespread use.  Research computing facilities often operate under tight budgets, with staff juggling multiple roles to support a diverse and demanding user base.  As a result, they have limited capacity to take on additional sustainability efforts.



We focus on the challenge of assessing the carbon footprint of scientific/HPC systems.  How can it be done comprehensively?  How can we create a basis for regular 
reporting?   A solution for it must have the following attributes:

\begin{itemize}
\item Achieve broad coverage, including the full range of  HPC systems.
\item Employ only limited data (a few data metrics) that are readily available
\item Require only modest human effort, as many HPC organizations are short-staffed
\item Provide a "gentle slope", allowing the addition of data metrics for unusual circumstances.
\end{itemize}

And we believe that, to be practicable, carbon footprint reporting for each system should require less than a person-hour of effort per year.

\section{Approach}

We assess the carbon footprint of HPC systems using the Top500 November 2024 list \cite{top500_2024} as a representative set of systems.\footnote{We expect to update for the June 2025 Top 500 list by the time of the workshop.}  Our approach builds on an existing tool for carbon-footprint modeling for computer systems called EasyC.   It  requires modest effort, low data, and provides high quality footprint reporting \cite{easyc_pearc}, see Figure \ref{fig:easyc_box}.



We begin with data from top500.org; however, this information is incomplete for many systems.  As many as 300-500 systems are missing specific pieces of required Top 500 reporting information (see Figure \ref{fig:structural-information}). To improve coverage and global applicability, we augment it with other publicly available sources (e.g., web page scraping) and refine the framework accordingly. Using these datasets, we conduct sensitivity analyses and assess the carbon footprint of the systems, as well as project future changes in the HPC carbon footprint.

Our results show that EasyC substantially increases the carbon-assessment coverage of these systems, reduces the data collection effort, and supports customization for diverse system configurations.









\section{Evaluating the Carbon Footprint of HPC}
\label{evaluation}

In the following sections, we assess the carbon emissions of HPC systems using the Top500 list as a proxy.   First, we apply a powerful carbon footprint modeling tool (EasyC
) to the available data from top500.org, and show how significant additional work is required to construct acceptable coverage of the 500 systems.
Second, we use the improved coverage to compute a high-quality assessment of the Top 500's carbon footprint.
Finally, we project the carbon footprint of the Top 500 for 2025-2030.


\subsection{Carbon Footprint of the Top500 HPC Systems}

We assess the carbon footprint of the Top 500 supercomputing systems in the world.  To our knowledge, this has never been done before. 
Thus, this study represents the first estimate of the global footprint of high-performance computing's carbon footprint.



\begin{figure}
    \centering
    \includegraphics[width=\linewidth]{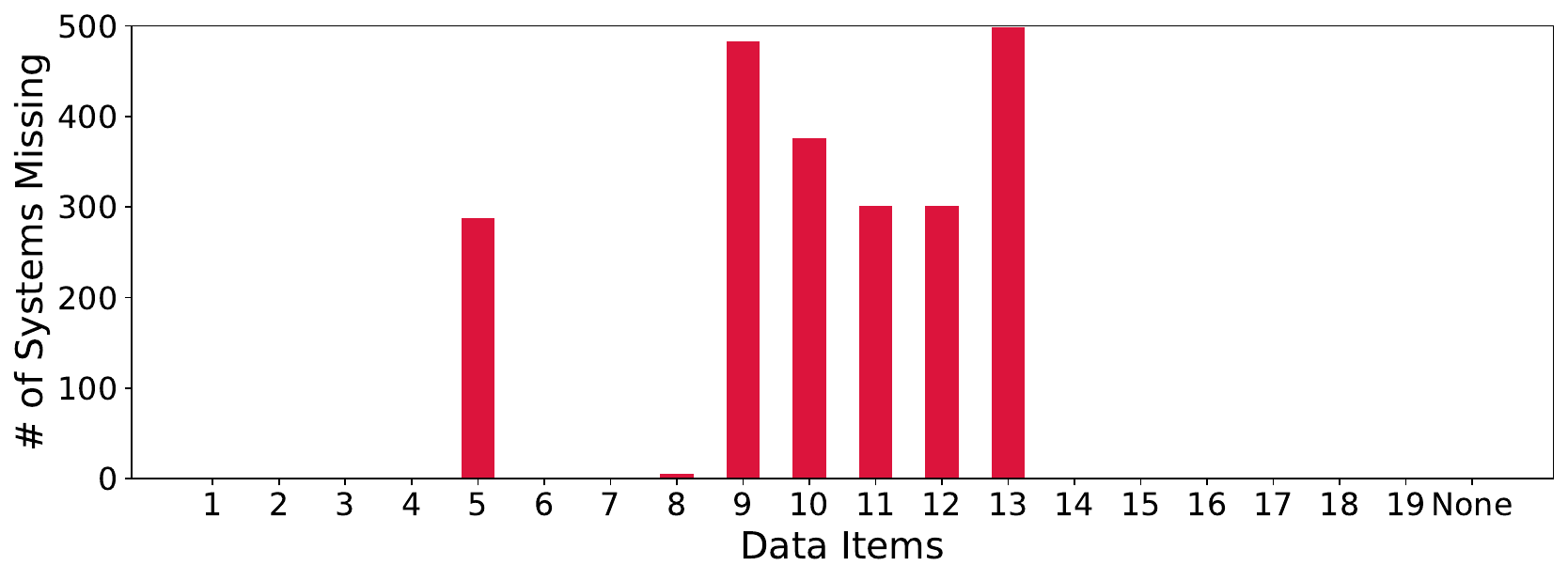}
    \caption{Structural information reported for different data items in the Top 500. Note: 'None' indicates that all information is reported.}
    \label{fig:structural-information}
\end{figure}

We use data from the Top 500 website. As part of its performance ranking, top500.org collects both performance and a variety of structural features of the systems (processor model, co-processor model, total cores, etc.)
as well as several performance attributes (
Rmax[TFlop/s], Rpeak[TFlop/s], Nmax, and others).
The structural and performance information is "high quality" for all 500 systems, because its a requirement for inclusion in the Top 500.  However there are data gaps that cause problems with carbon-footprint assessment (see Figure \ref{fig:structural-information}).

\begin{figure}[t]
\noindent\fbox{%
    \parbox{\linewidth}{%
    \includegraphics[width=0.95\linewidth]{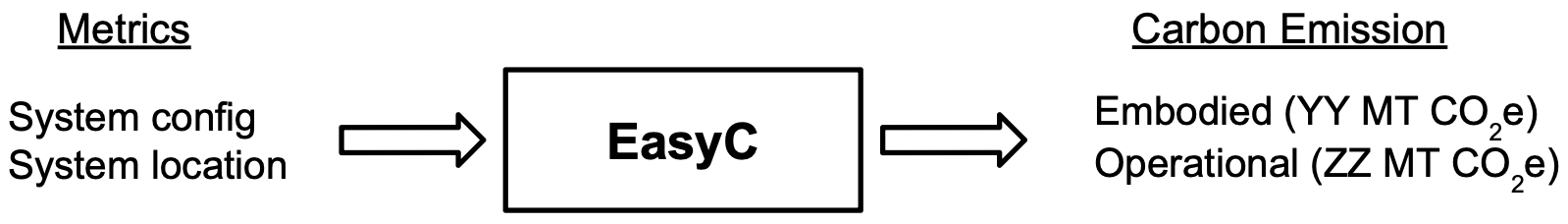}
    \textbf{EasyC}: The EasyC tool calculates the carbon footprint (operational and embodied), using sophisticated models that identify the key data metrics required.  It needs just 7 key data metrics\cite{easyc_pearc}.  This differs from the widely used GHG Protocol that can require hundreds of metrics.
    }%
}\\
\caption{EasyC Carbon Footprint Tool} 
    \label{fig:easyc_box}
\end{figure}

\begin{figure}[h!]
    \centering
    \begin{subfigure}{0.5\textwidth} 
        \includegraphics[width=1\textwidth]{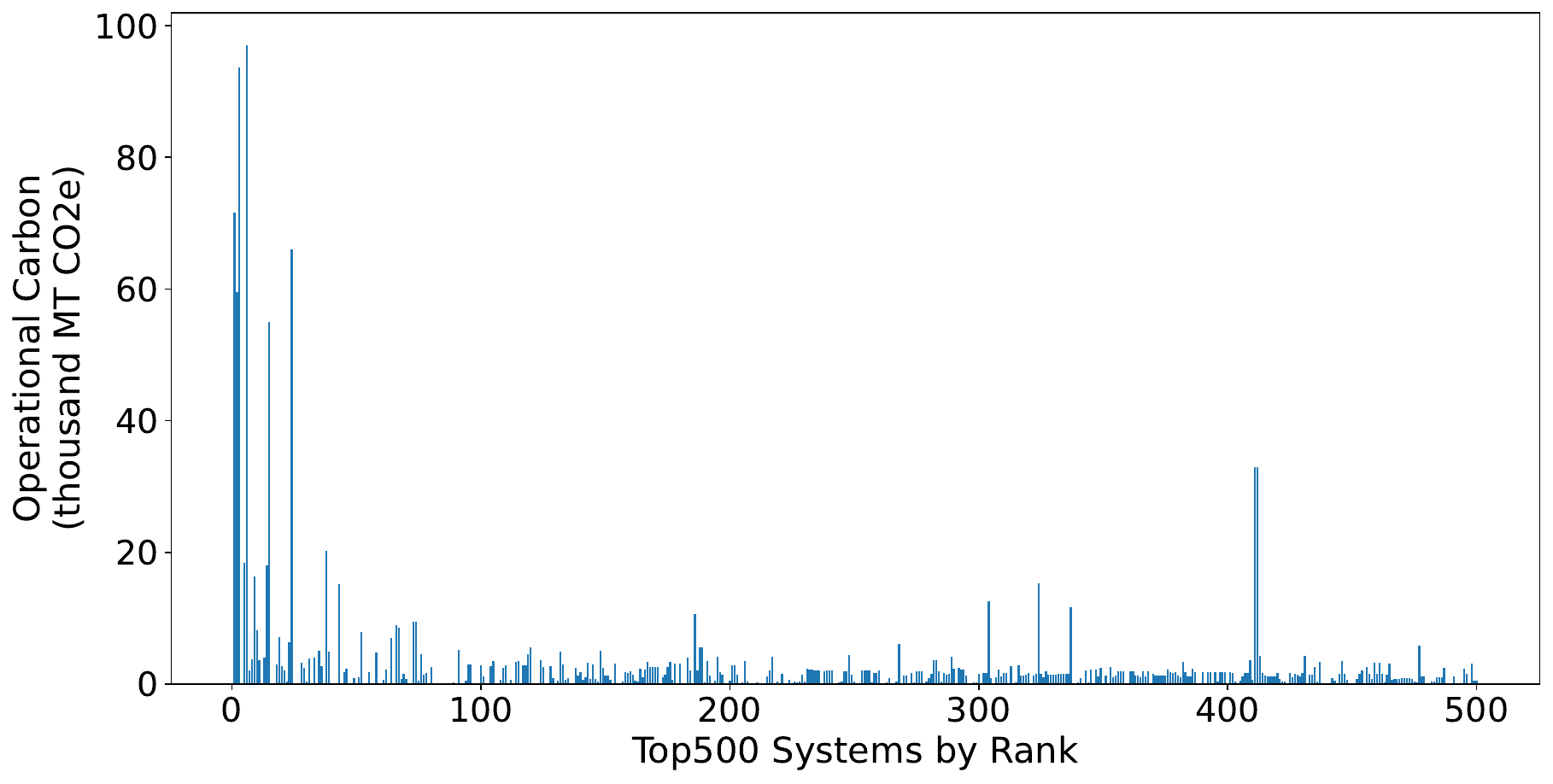}
         \caption{Operational Carbon}
         \label{fig:operational_carbon_rank}
    \end{subfigure}
    \hfill
    \begin{subfigure}{0.5\textwidth} 
        \includegraphics[width=1\textwidth]{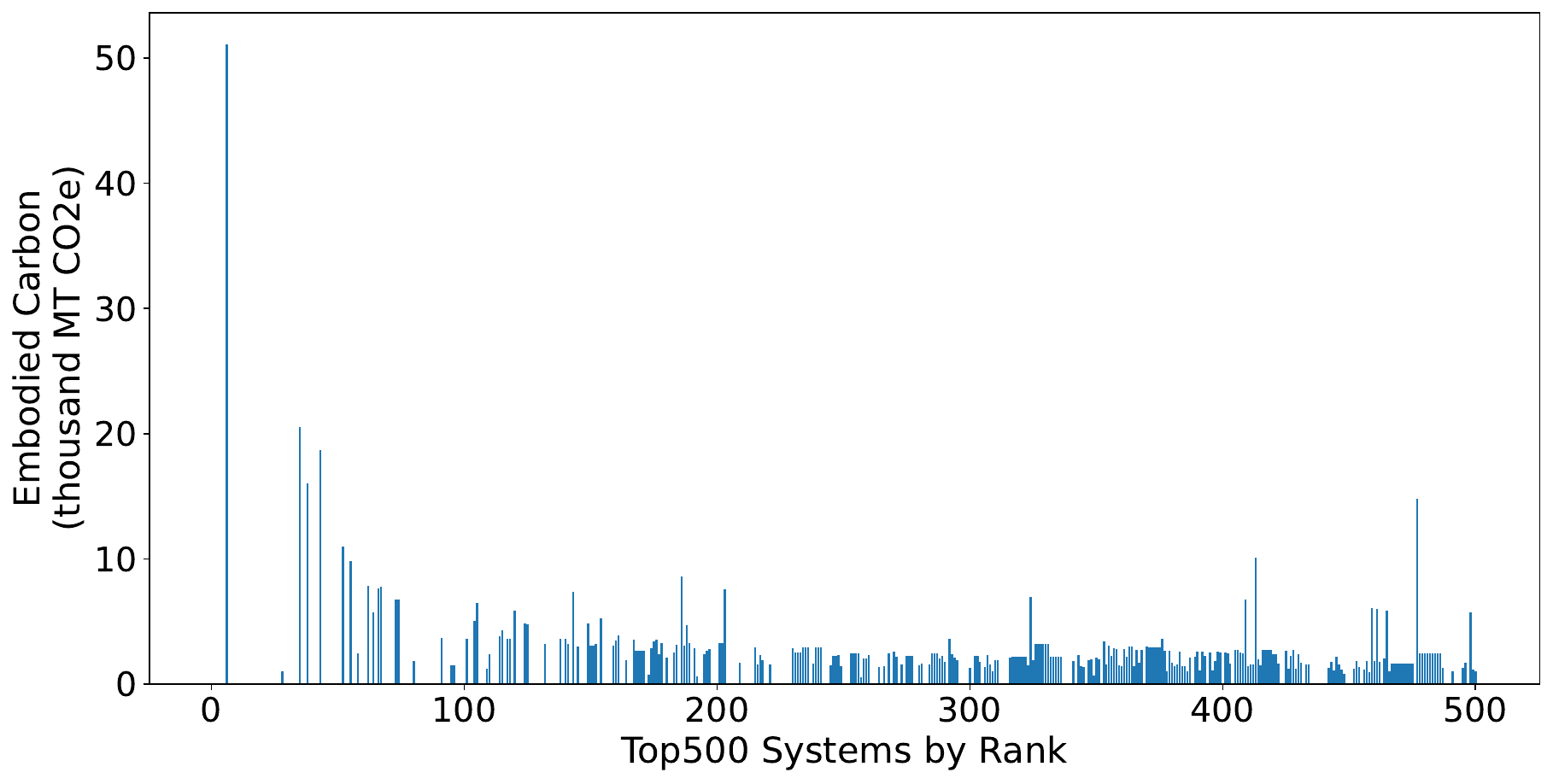}
        \caption{\label{fig:embodied_carbon_rank} Embodied Carbon}
    \end{subfigure}
    \caption{Top 500 Carbon Footprint vs Rank (Top500.org data)} 
    \label{fig:carbon_rank}
\end{figure}

Based on the available Top 500 structural information, we use the EasyC tool 
to assess the carbon footprint for all the systems that sufficient data is available.  This is 391/500 systems for operational carbon and 283/500 for embodied carbon.  Figures \ref{fig:operational_carbon_rank} and \ref{fig:embodied_carbon_rank} shows operational (1 Year) and embodied carbon versus rank.  This coverage, while the best to date, is still too poor to construct an overall Top 500 assessment because it covers little more than 
half of the systems.
Further, the systems present and missing are highly skewed; the highest-performing systems, the largest contributors, are disproportionately missing.

\noindent \textbf{Steps to Increase Coverage} 
How to increase coverage?  Interpolation is not robust with such a sparse collection.  We turn to the core idea of EasyC,  
which is that intelligent use of just a few key data metrics can be used to create accurate carbon modeling (see Figure \ref{fig:easyc_box}).  It's natural to ask how and what additional sources of data could be exploited to enable better coverage. 
Specifically, we consider two data metric scenarios:
(1) Baseline, uses only data available on Top500.org, and (2) Baseline+Public Info, supplements the data with publicly available information on other web sites.
We first assess coverage, and then quantitative carbon metrics.

{\small 
\begin{table}[ht]
\caption{Data EasyC requires, but not 
available on Top500.org, public disclosure}
\label{tab:data-missing}
\centering
\begin{tabular}{|p{70pt}|r|r|}
\hline
 Type & \# Systems Incomplete & \# Systems Incomplete \\
      & [Top500.org] & [Other Public] \\[0.5ex]
\hline \hline
Operation Year & 0 & 0 \\ \hline
\# of Compute Nodes & 209 & 86 \\ \hline
\# of GPUs & 209 & 86 \\ \hline
\# of CPUs & 0 & 0 \\ \hline
Memory Capacity & 499 & 292 \\ \hline
Memory Type & 500 & 292 \\ \hline
SSD Capacity & 500 & 450 \\ \hline
System Util (opt.) & 500 & 497\\ \hline
Annual Power Consumed (opt.) & 500 & 492 \\ \hline \hline
\# Data Metrics Missing & 9 & 9 \\ \hline
\end{tabular}
\label{table:top500_data}
\end{table}
}

Because of their leading edge nature, the Top 500 systems are particularly challenging to analyze. Even the minimum data required to estimate carbon emissions is missing for many systems. Table \ref{table:top500_data} shows the extent of data incompleteness when considering both top500.org and other publicly available sources. 
To compute
operational carbon requires actual power consumption and power source. Embodied carbon requires system configuration and component information.
Exacerbating this, some systems use early or unique compute devices (eg MI300A, Fugaku A64FX, Sunway SW26010, etc.)  particularly for embodied carbon.  Essentially all of the Top 500 do not report actual power consumption.  Many of the systems are commercially operated or government systems with no public information available.  Only a few are "open science" systems with good configuration information available.

Further, we found that none of the systems provided reporting under the GHG protocol (so there are no reports to compare with).  


\begin{figure}[ht]
    \centering
    \begin{subfigure}{0.24\textwidth} 
        \includegraphics[width=1\textwidth]{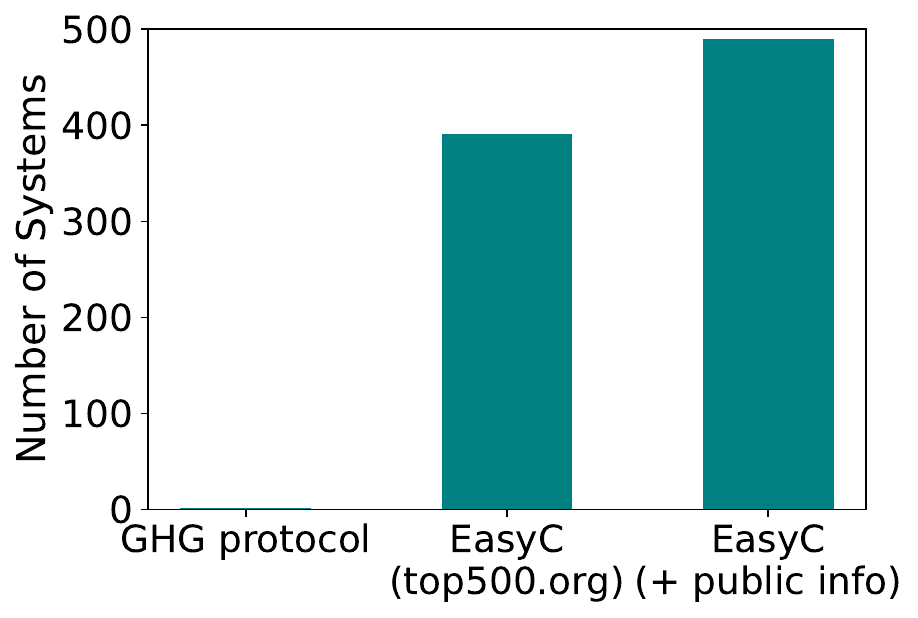}
         \caption{Operational Carbon}
    \end{subfigure}
    \begin{subfigure}{0.24\textwidth} 
        \includegraphics[width=1\textwidth]{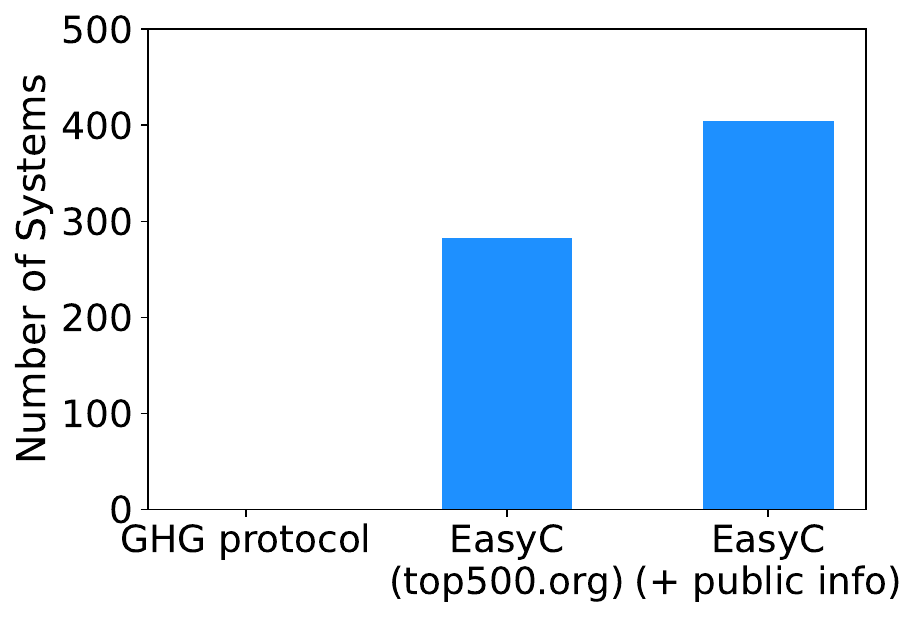}
        \caption{\label{fig:easyc_coverage_embodied} Embodied Carbon}
    \end{subfigure}
    \caption{Carbon Footprint Reporting Coverage} 
    \label{fig:easyc_coverage}
\end{figure}

In Figure \ref{fig:easyc_coverage}, we show how the coverage of systems increases varies with different data metric availability.  Using the GHG detailed carbon accounting method, few of the Top 500 systems reports operational and NONE report embodied. As we add public data, coverage of the Top 500 operational carbon increases from 
78\% of systems on top500.org using only the data available there. Including publicly available data (number of compute nodes, number of GPUs, number of CPUs, memory capacity, and SSD capacity), this coverage jumps to 98\%, allowing us to estimate almost all systems' operational emissions. 
For embodied carbon emissions, we see a 1.43x improvement in coverage for estimating embodied carbon emissions. This is because data crucial for embodied carbon estimates like the number of nodes, accelerators, or processors is often missing from top500.org. 


\begin{figure}[ht]
    \centering
    \begin{subfigure}{0.5\textwidth} 
        \includegraphics[width=1\textwidth]{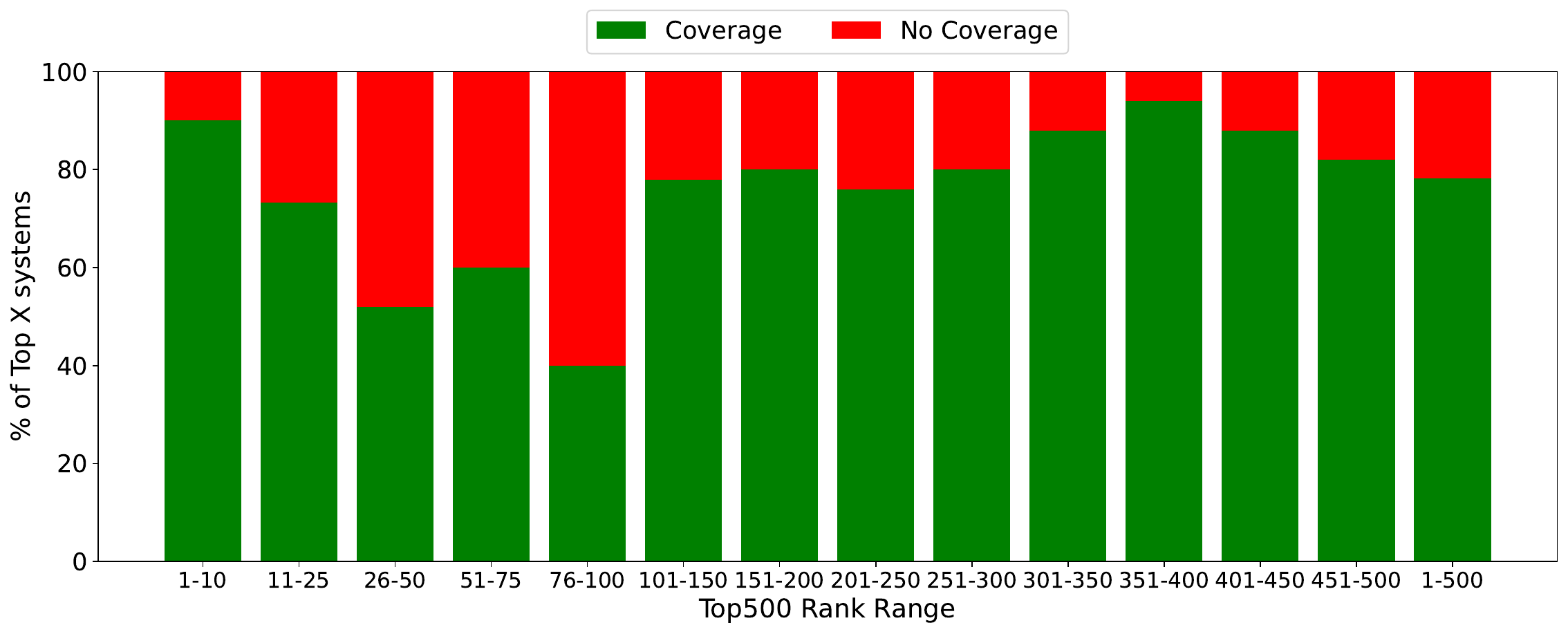}
         \caption{With Top500.org data}
    \end{subfigure}
    \hfill
    \begin{subfigure}{0.5\textwidth} 
        \includegraphics[width=1\textwidth]{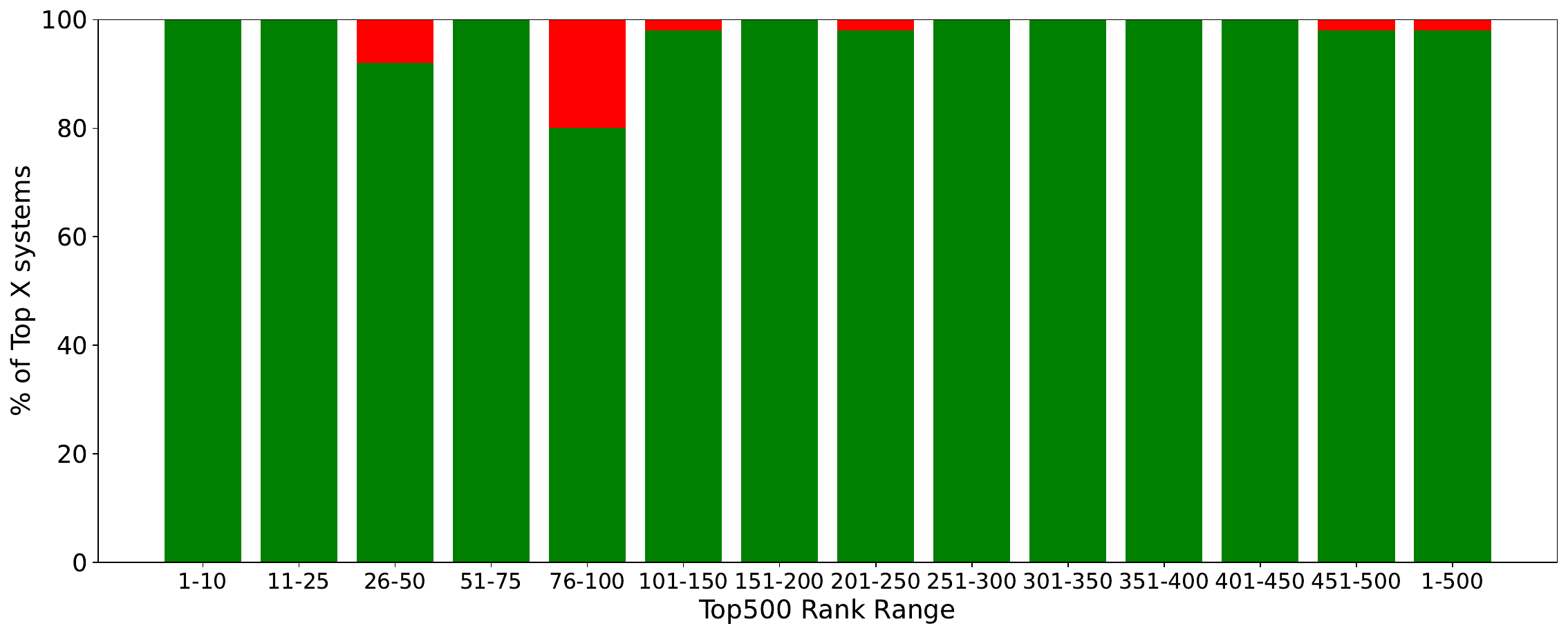}
        \caption{\label{fig:easyc_top500_public} With Top500.org and other public data}
    \end{subfigure}
    \caption{Operational carbon coverage of EasyC for varied system ranks (two data input scenarios).} 
    \label{fig:easyc_top500_range_operational}
\end{figure}

\begin{figure}[ht]
    \begin{subfigure}{0.5\textwidth} 
        \includegraphics[width=1\textwidth]{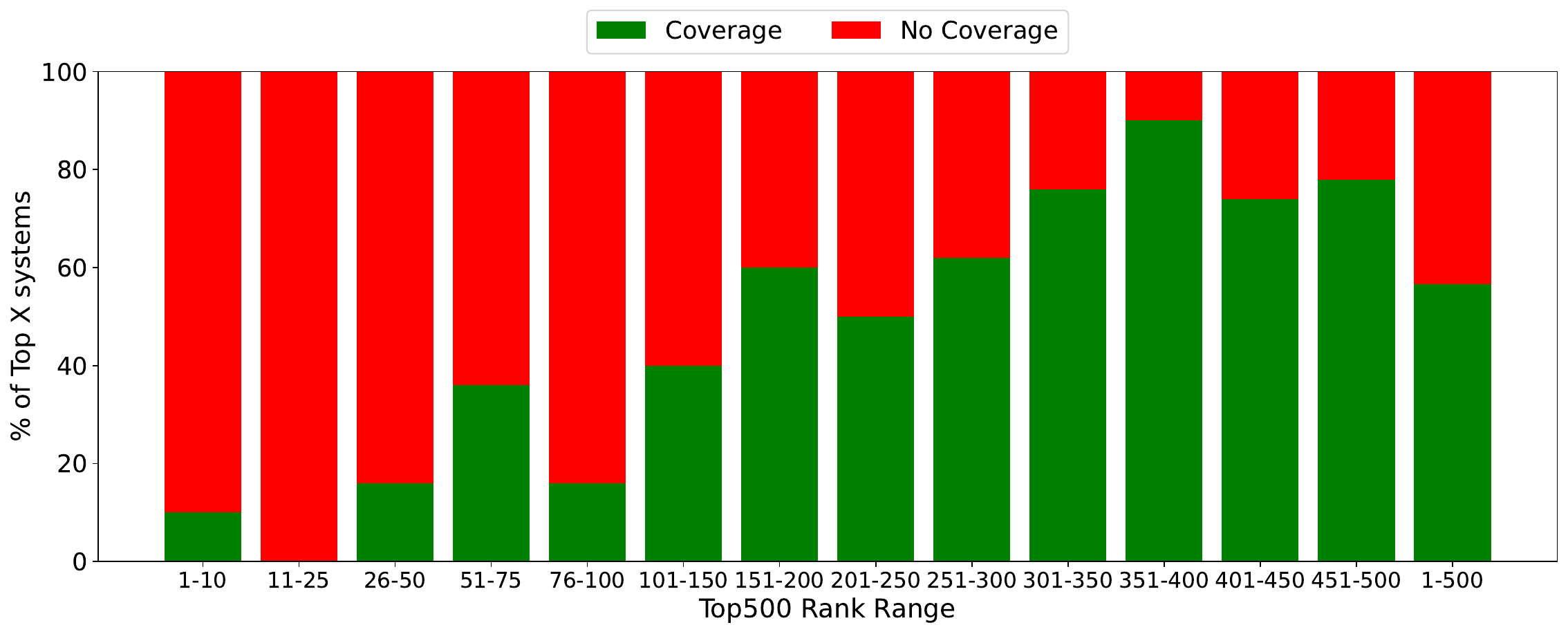}
         \caption{With Top500.org data}
    \end{subfigure}
    \hfill
    \begin{subfigure}{0.5\textwidth} 
        \includegraphics[width=1\textwidth]{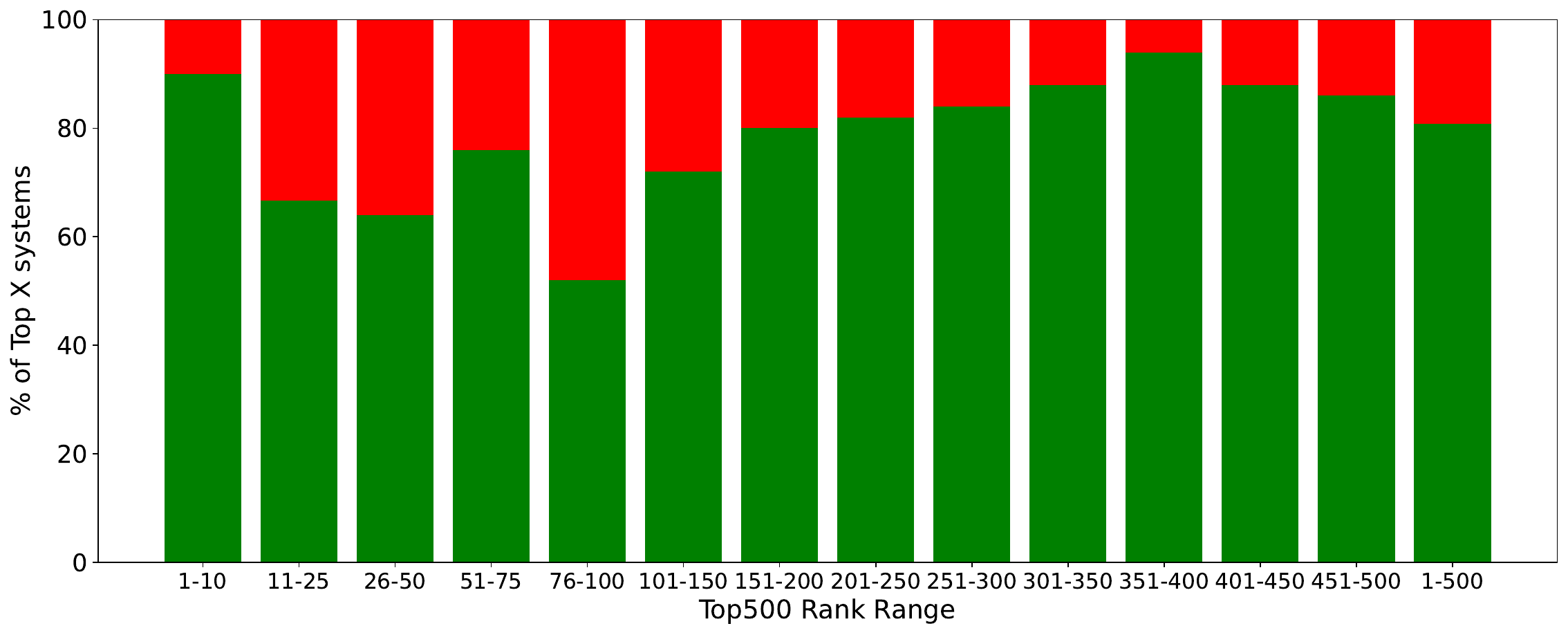}
        \caption{\label{fig:easyc_top500_public} With Top500.org and Public data}
    \end{subfigure}
    \caption{Embodied carbon coverage of EasyC  for varied system ranks (two data input scenarios).} 
    \label{fig:easyc_top500_range_embodied}
\end{figure}


Figures \ref{fig:easyc_top500_range_operational} and \ref{fig:easyc_top500_range_embodied} present the varying coverage across Top500 system ranking ranges. 
The operational carbon emissions of many Top 500 systems can be estimated using data from top500.org.
 Significant gaps emerge surprisingly high in the rankings 26-50, 51-75, and 76-100. The main reason is that many systems do not report their power consumption, which is essential when information on the number of compute nodes and GPU nodes is unavailable. The addition of public information fills these gaps, rendering nearly full coverage. 

For embodied carbon, the situation is much worse.  For many systems in the Top 150, there was insufficient data to make an estimate, even with the limited data requirements of EasyC. 
The major reason for this is that top systems today make heavy use of an increasingly diverse set of accelerators (e.g., Nvidia, AMD, many versions).  Top500.org does not capture adequate accelerator information.
For most of the systems 151-500 systems, they are CPU-based.  Because they do not have accelerators, the number of CPU cores per node and total CPU cores that are captured at top500.org are sufficient for embodied carbon.
We can see in Figure \ref{fig:easyc_top500_range_embodied}b that incorporating additional publicly available data on which accelerators were used is essential to improve coverage. 

\subsection{Constructing an Estimate of the Full Top 500}

With much higher coverage, we employ simple statistical techniques to construct an assessment of carbon footprint for the full Top 500 supercomputers.

\begin{figure}[h!]
    \centering
    \begin{subfigure}{0.4\textwidth} 
        \includegraphics[width=0.9\textwidth]{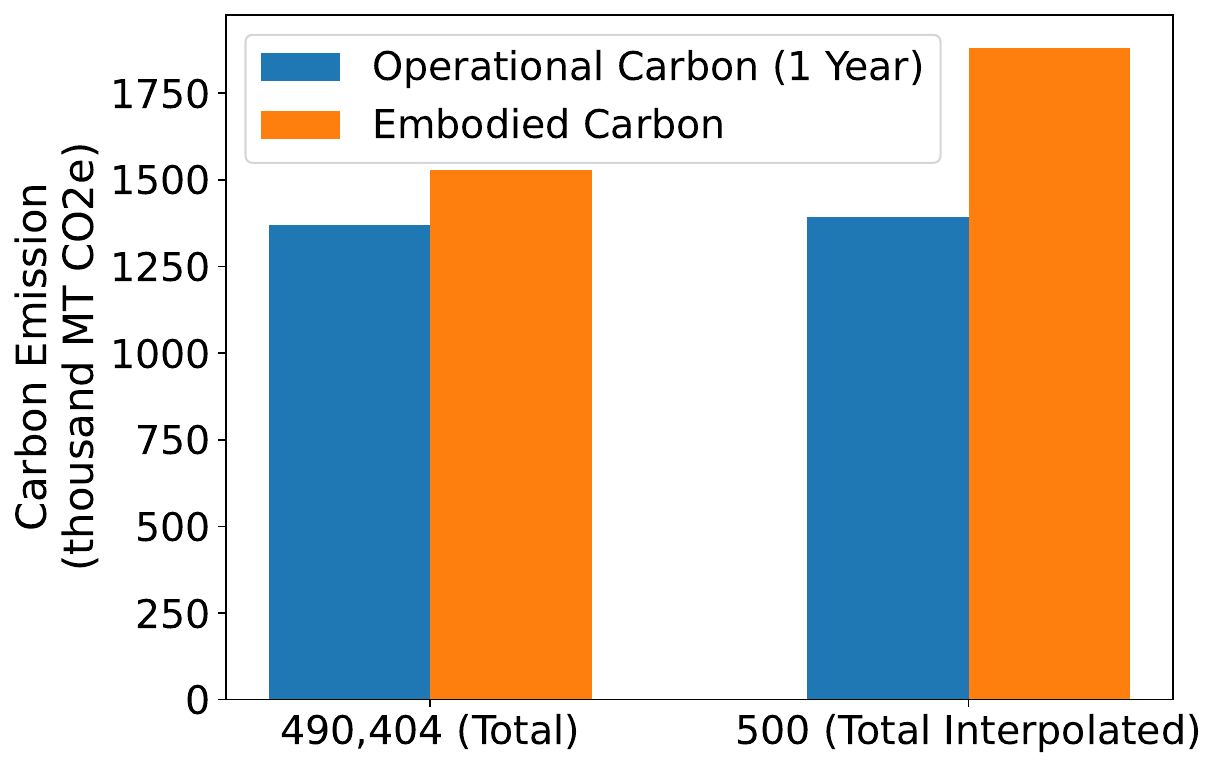}
         \caption{Total Carbon Footprint}
         \label{fig:top500-carbon-footprint_total}
    \end{subfigure}
    \hfill
    \begin{subfigure}{0.4\textwidth} 
        \includegraphics[width=0.9\textwidth]{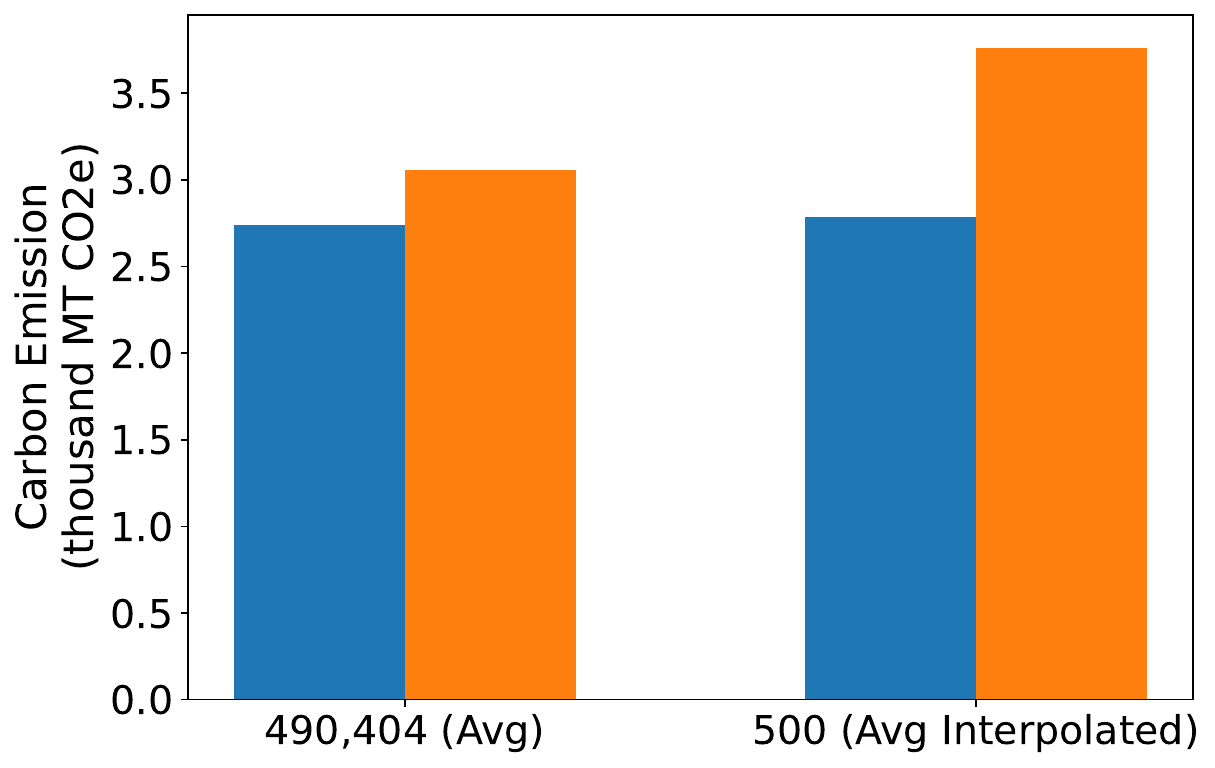}
        \caption{\label{fig:top500-carbon-footprint_avg} Average System Carbon Footprint}
    \end{subfigure}
    \caption{Total Operational carbon (1 year) and Embodied Carbon for systems covered by Top500 + other public data  and then full Top 500, interpolating for the missing systems.  Then average carbon for the same sets of systems.} 
    \label{fig:top500-carbon-footprint}
\end{figure}

To fill the gaps of 100's of systems,  we interpolate the carbon footprint for the systems missing data using  the average of the nearest 10 peers (5 lower and 5 higher) in the Top 500.\footnote{If the peers are also incomplete, we use the next closest peers.}  The data in Figure \ref{fig:top500-carbon-footprint}a represent the carbon footprint of HPC.  On the left, we present the EasyC carbon footprint for all the systems covered by Top500.org plus other public information.  On the right, we include those and add the interpolated systems. Continuing, we also compute the average operational and embodied carbon of systems in the systems with enough data (for EasyC) and for the full 500, completed by interpolation.  This brings the Operational carbon for the Top 500 to 1.39 million MT CO2e, adding the missing 10 systems increased operational footprint by only 1.74\%.

\begin{figure}[h!]
    \centering
    \begin{subfigure}{0.5\textwidth} 
        \includegraphics[width=1\textwidth]{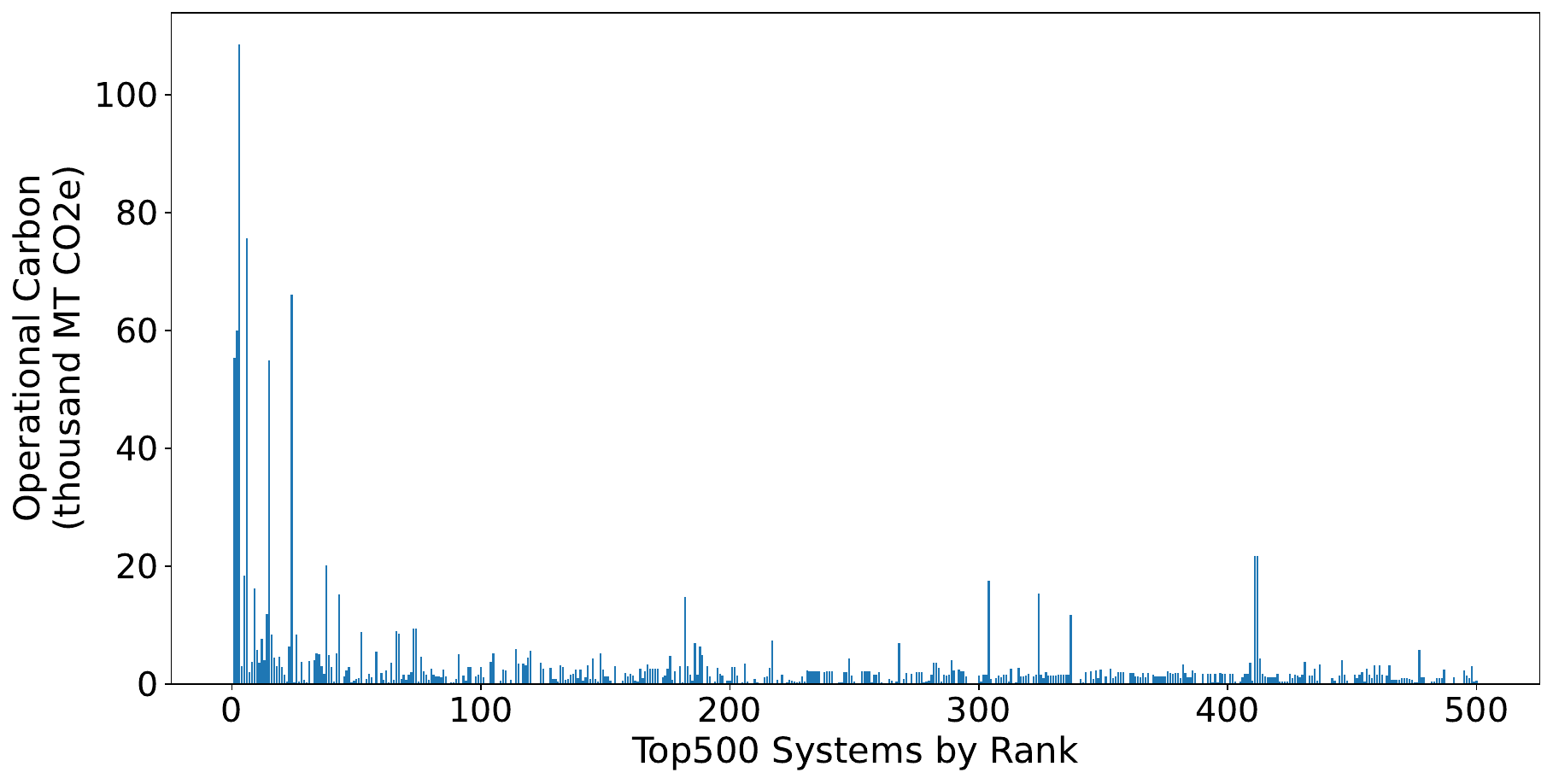}
         \caption{Operational Carbon}
         \label{fig:operational_carbon_complete}
    \end{subfigure}
    \hfill
    \begin{subfigure}{0.5\textwidth} 
        \includegraphics[width=1\textwidth]{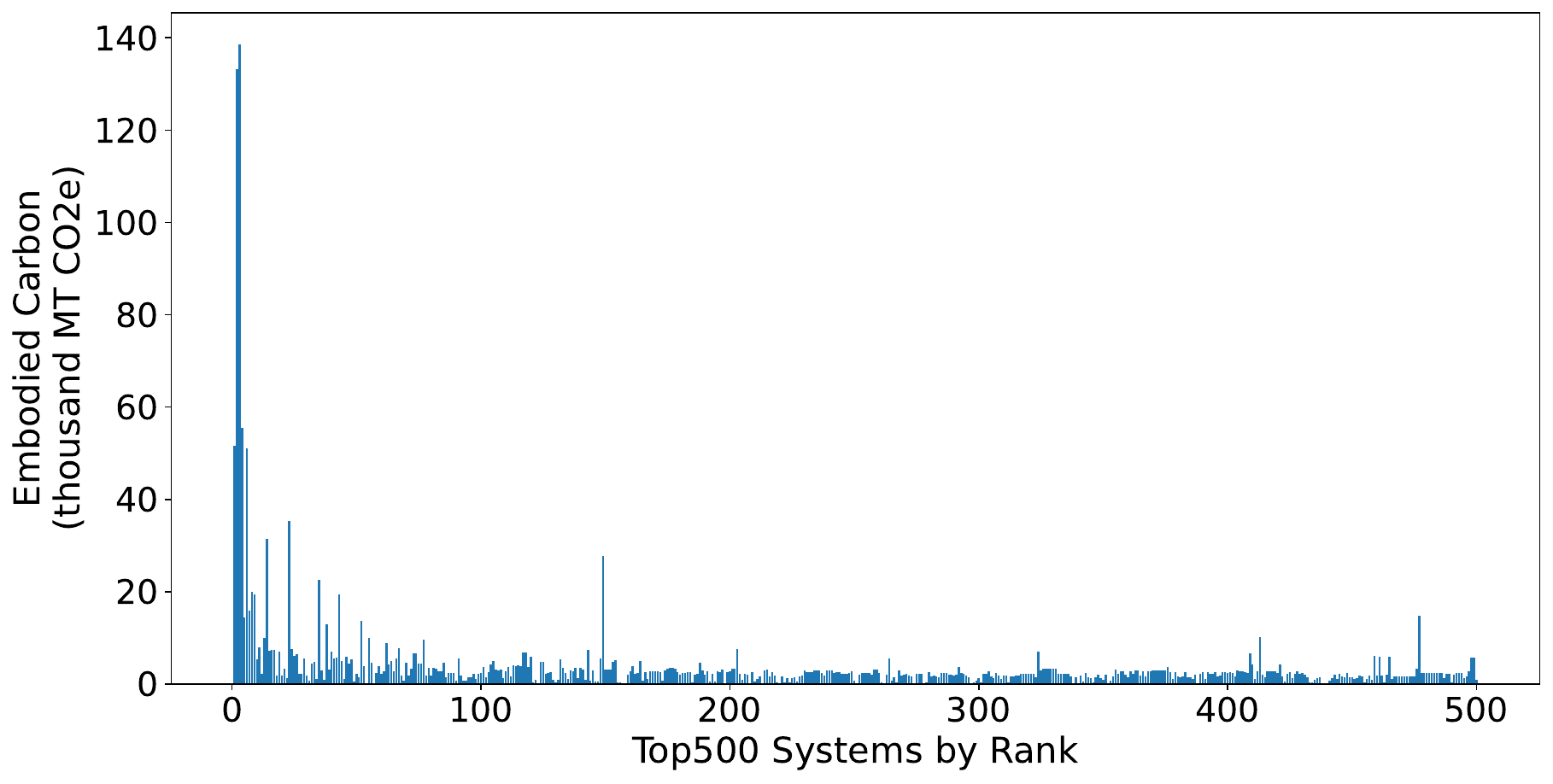}
        \caption{\label{fig:embodied_carbon_complete} Embodied Carbon}
    \end{subfigure}
    \caption{Top 500 Systems Carbon Footprint, by rank.  Full assessment, includes data generated by EasyC on top500.org and other public web data, augmented by interpolation.} 
    \label{fig:top500-carbon-complete}
\end{figure}

Adding the missing 96 systems increased embodied carbon to 1.88 million MT CO2e, an increase of 23.18\%.  This operational carbon is equal to one year's emissions for 325,000 gasoline-powered vehicles or 3.5 billion vehicle miles.  The embodied carbon is equivalent to one year's emissions for 439,000 gasoline-powered vehicles or 4.8 billion passenger miles.  Of course the embodied carbon is 1-time for the lifetime of the computer system so it would be smaller if annualized.
In Figure \ref{fig:top500-carbon-complete}b, the average operational and embodied carbon for Top 500 systems is presented. Each is thousands of MT CO2e, comparable to that of thousands of homes.







In Figure \ref{fig:top500-carbon-footprint}, 
we report the aggregate carbon footprint for the Top 500, both with the systems for which we have sufficient data. 
For 490 systems, the total operational carbon is 1.37 Million MT CO2e, and for 404 systems the total embodied carbon is 1.53 Million MT CO2e.  


\begin{figure}
    \centering
    \includegraphics[width=0.9\linewidth]{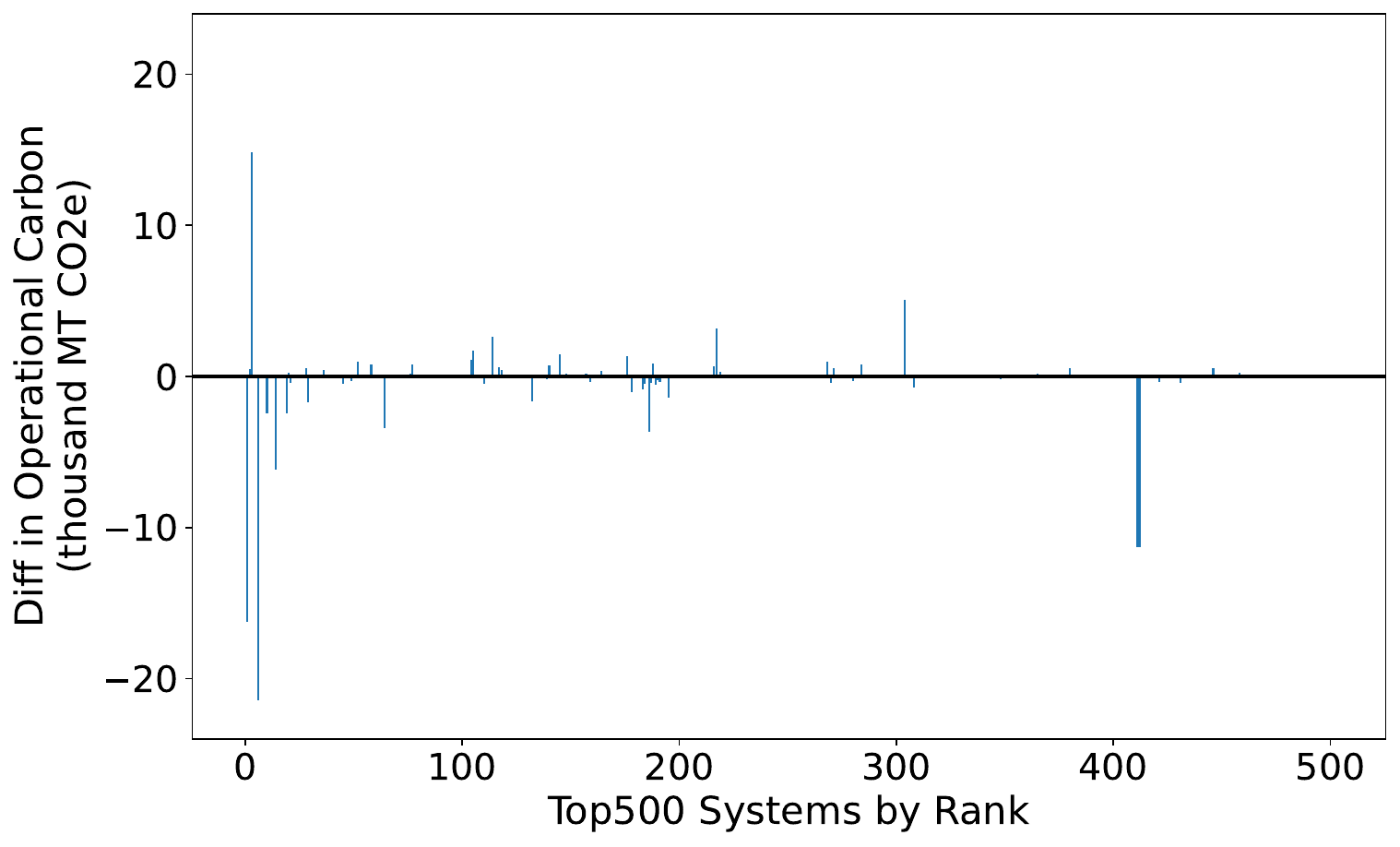}  
    \includegraphics[width=0.9\linewidth]{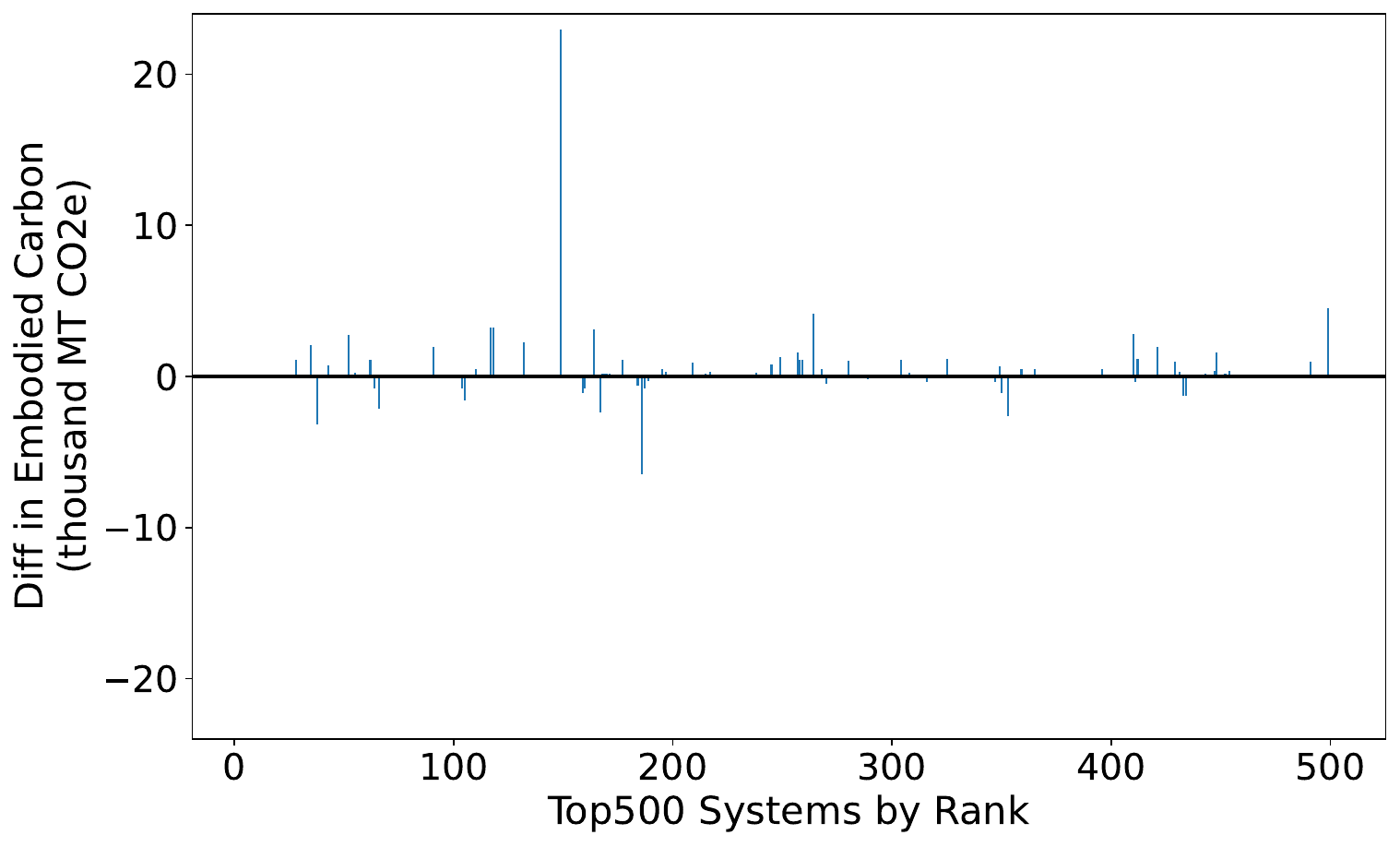}

    
    \caption{How Operational and Embodied carbon footprint changes between Baseline (Top 500 data) and Baseline+PublicInfo.}
    \label{fig:embodied-difference-info}
    \label{fig:operational-difference-info}
\end{figure}

The complete Top 500 results for each system are shown in Figure \ref{fig:top500-carbon-complete}. It includes all of the data generated by EasyC on top500.org and public web data, augmented by 10 systems for operational and 96 systems for embodied modeled by interpolation.  In short, it includes a model for all 500 systems in the Top 500 list.

As a sensitivity study, we show the difference that adding public information made 
in Figure \ref{fig:embodied-difference-info}. 
For operational carbon, we can see that adding public information does not change the overall results significantly.  For operational carbon, we see refinement to a narrower region average carbon intensity (ACI) of power, which can increase or decrease by as much as  77.5\%.  But even a number of these do create a large change, the total change for the entire Top 500 is only 2.85\% (38 thousand MT CO2e). 
For embodied carbon,  there are larger changes, mostly increasing the carbon footprint, but these have small contribution.  The biggest change is due to large systems where no estimate was previously possible (not shown).  Overall this produces and increase of 670.48 thousand MT CO2e, for an 78\% change overall. The use of novel accelerators, not documented in public information, is the largest problem.  Approximating these accelerators with mainstream GPUs  produces systematic underestimates of silicon size.


Overall, our results show that  operational carbon emissions can typically be derived directly from top500.org.  This is true across the range from 1-500, and for systems with and without accelerators.  Embodied carbon is more difficult, as systems with accelerators are difficult due to their diversity, and growing number of types. But with additional other public data, our results show that EasyC can  compute carbon emissions of almost all the systems on Top500, and with little effort.

\excise{\subsection{Drilling Down on EasyC Accuracy}

\aac{accuracy discussion}

\aac{why a full top500 study may NEVER be possible with the detailed approach}



Detailed methodologies have extensive data requirements and require immense effort. The configurations of the systems change frequently due to hardware failure, hardware replacement.   
The lack of reliable data makes it nearly impossible to evaluate accuracy across the entire Top500 list. However, we compare EasyC's results to the systems analyzed in the state-of-the-art study by Li et al. \cite{li2023sustainablehpc}.


\begin{figure}[h!]
    \centering
    \includegraphics[width=0.9\linewidth]{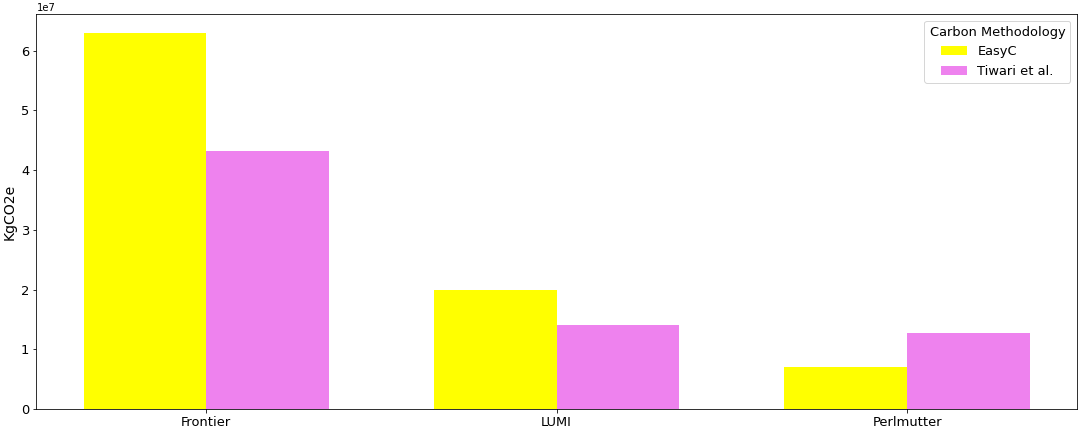}
    \caption{Comparing EasyC's embodied carbon estimates with Li et al. (Not including storage component) }
    \label{fig:easyc_tiwari}
\end{figure}

We evaluate for the embodied carbon data of the following HPC systems: Frontier, LUMI, and Perlmutter. Figure \ref{fig:easyc_tiwari} shows the embodied carbon computed by EasyC and Li et al. We don't include the storage component for comparison because Li et al. consider the entire systems storage capacity not just the storage on the compute nodes. We find that computed values can vary in the range of 40-45.71\% error. This due to various discrepancies in data like carbon emission values, die area size, etc. This shows that the accuracy for most of these systems by EasyC is in the range of the errors by the detailed methodologies.
}

\subsection{How Carbon Footprint will Change for the Top 500?}

\begin{figure}[ht]
    \centering
    \begin{subfigure}{0.45\textwidth} 
        \includegraphics[width=1\textwidth]{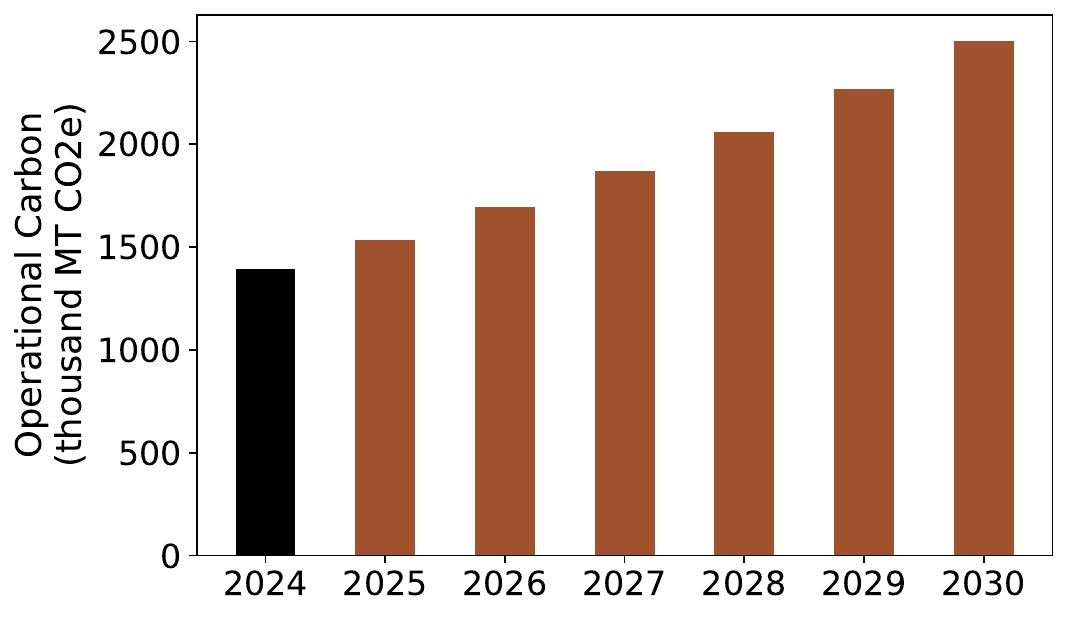}
         \caption{Operational Carbon}
    \end{subfigure}
    \hfill
    \begin{subfigure}{0.45\textwidth} 
        \includegraphics[width=1\textwidth]{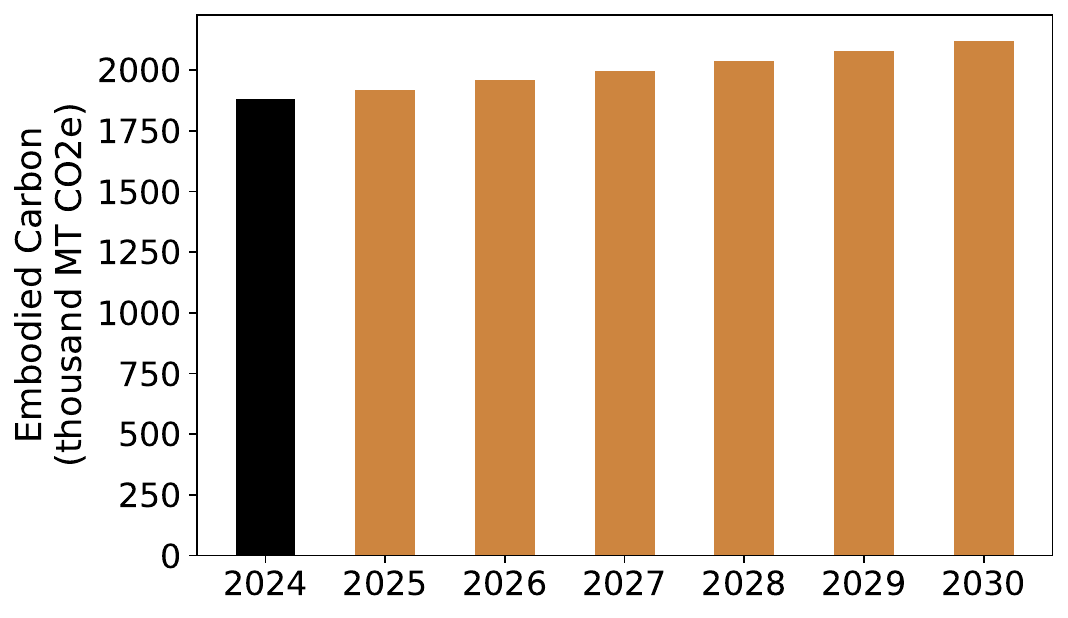}
        \caption{\label{fig:embodied_projection} Embodied Carbon}
    \end{subfigure}
    \caption{Projected Top 500 Operational and Embodied Carbon (2025-2030)} 
    \label{fig:projected-carbon}
\end{figure}

We project the carbon footprint of the Top 500 HPC systems for 5 years. An average of 48 systems was added to each new list in each cycle, over the past two years.  With this turnover comes a 5\% increase in operational carbon, and 1\% increase in embodied.  Annualized, this is 10.3\% growth in operational and 2\% growth in embodied carbon.


We project carbon footprint of the Top 500 based on these growth rates (see  Figure \ref{fig:projected-carbon}).
Operational carbon's 10.3\% annual growth reflects both increasing  computing power represented in the Top 500 systems, but also the slowing benefits of Moore's Law and the reality of post-Dennard scaling, as power consumption continues to increase rapidly.  By 2030, Top 500's operational carbon is nearly double that of 2024.  Clearly architectural customization and accelerators is not enough to reduce power consumption.

In comparison, the embodied carbon emissions is growing slower (only 1.02x or 2\% per year), suggesting its a lesser concern.  However, we would advocate caution as embodied carbon is heavily influenced by storage system -- not a focus of this study.



\begin{figure}[ht]
    \centering
    \begin{subfigure}{0.4\textwidth} 
        \includegraphics[width=1\textwidth]{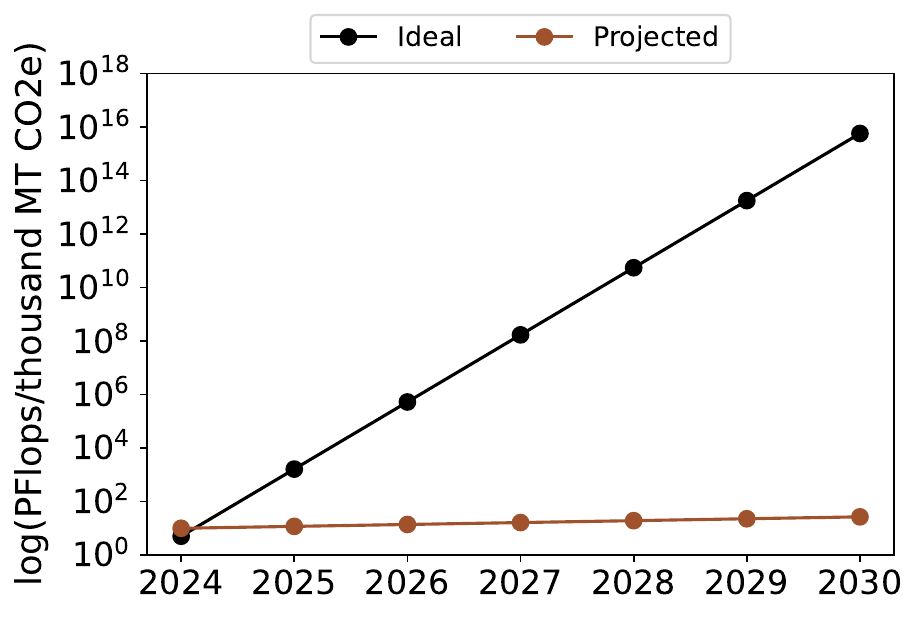}
         \caption{Operational Carbon}
    \end{subfigure}
    \hfill
    \begin{subfigure}{0.4\textwidth} 
        \includegraphics[width=1\textwidth]{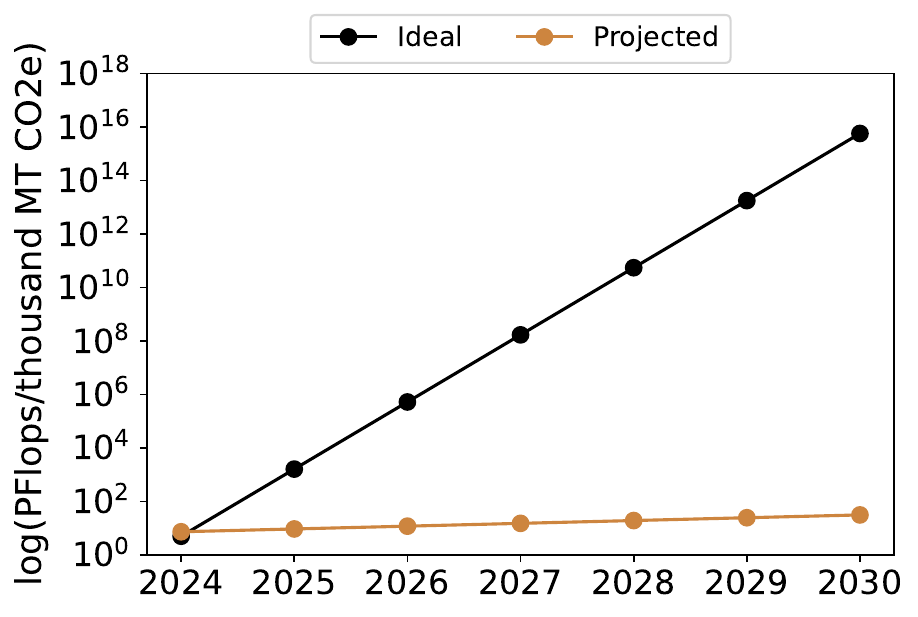}
        \caption{\label{fig:embodied_projection_perf} Embodied Carbon}
    \end{subfigure}
    \caption{Projected Top 500 Operational and Embodied Performance to Carbon Ratio (2025-2030)} 
    \label{fig:projected-carbon-perf}
\end{figure}





Compute scaling has benefitted from dramatic increases on performance per unit (size, power, etc.) over more than seven decades \cite{moores_law,dennard_scaling}.  In recent times, those microelectronics scaling drivers have slowed or even stopped.  With a sustainability focus, we consider performance per carbon footprint for the Top 500 HPC systems (see Figure \ref{fig:projected-carbon-perf}).

The projection shows a continued increase in perf/carbon, but at an annual rate of 0.2 PFlop/s per thousand MT CO2e.  While there is still progress, this is dramatically slower than what we experienced for decades under Dennard scaling, and parallelism scaling where we'd expect 2x performance / unit power every 18-month generation (plotted for comparison).
Note that the current increase in performance / unit carbon is not sufficient to compensate for the rapid growth in the use of computing.  As the net growth of total carbon footprint of the Top 500 continues at  10.3\%/annum (see Figure \ref{fig:projected-carbon}).


\section{Discussion and Related Work}
\label{related_work} 


\subsection{Organizational Reporting (GHG Protocol)}


Many organizations including Google, Microsoft, TSMC, Intel, United Airlines, Emirates, Harvard University, Argonne National Laboratory, and others publish annual sustainability reports\cite{google_sustainability_2024, microsoft_sustainability_2024, tsmc_2024, intel_environment_report2024, united_environment_report2024, emirates_environment_report2024, harvard_environment_report2023, ANL_environment_report2023} based on the GHG Protocol\cite{ghg_protocol2024}. These reports take a broad view, covering carbon emissions from construction, employee commuting, purchased goods and services, capital goods, IT, and more. However, their scope is generally at the organizational level rather than the level of individual computing systems, making it difficult to directly assess the carbon footprint of specific computing systems which is the focus of this work.







\subsection{Studies/Reporting of Individual Computer Systems}

Several recent academic studies have estimated the carbon emissions of specific systems \cite{li2025ecoserve,li2023sustainablehpc,benhari2025top500}
However, their methodologies are based on the GHG Protocol, which requires extensive data inputs and significant effort for each system. Thus, their approaches cannot be easily adapted to assessing a wide range of computing systems, the focus of this paper.

\subsection{GHG Protocol vs EasyC}
The GHG Protocol \cite{ghg_protocol2024} is an accepted standard for carbon accounting.  It is based on comprehensive data collection and tabulation.
This approach means that accuracy is dependent on every piece of data needed, a classical accounting approach.  Further, exhaustive data collection and tabulation is a major effort, and invites the include of inaccurate data.

Research computing systems are complex aggregations of thousands of devices, each comprised of hundreds of components, produced by complex supply chains.  As a result, each device and component can have important differences in life-cycle history and performance (e.g. power efficiency).
For example, a system may use DRAMs manufactured in South Korea, Japan, and the United States. Since each manufacturing site differs, even DRAM with the same capacity can vary in components and materials used.
Further, even if possible to obtain data for each component and device, each inclusion incorporates additional data inaccuracies into the larger computation.  There's no reason to believe that "statistical variation" will reduce error, as these inputs are rife with systematic error.
For example, errors can arise from using outdated or inaccurate carbon intensity data for electricity, often due to limited public availability, inconsistent time granularity, or omission of on-site energy sources.

Finally, the effort required to do a diligent GHG Protocol computation for a computer system is daunting, perhaps requires weeks of effort, and thus may be impractical for many research computing facilities that are often staffing-limited.

However, EasyC uses a statistical approach to identify the key metrics which are predictive of carbon footprint. 
As a result, EasyC reduces the data metrics needed to only seven, dramatically lowering effort required.  This enables internal assessment of research computing systems, and also external assessment of groups of computing systems such as the Top 500.

\section{Summary and Future Work}
Detailed carbon emission computation methodologies fail to achieve high accuracy even with comprehensive data and significant effort. This makes it impossible for research computing facilities with limited resources to use it. 
However, our evaluation on Top500 systems is a first large-scale assessment of HPC's carbon footprint. The fact that it was even feasible demonstrates EasyC's ability to compute carbon emissions for diverse heterogeneous systems, unlike more detailed approaches based on GHG protocol and life-cycle assessment.  

Going forward, we hope that this work will lead to widespread carbon emission reporting of computing systems.  For example, we would like to model carbon footprint for all of the US National Science Foundation ACCESS scientific computing sites, those of the US Department of Energy, 
or of similar such systems in Europe or China.


\section*{Acknowledgment}
We thank the anonymous reviewers for the insightful reviews. This work is supported in part by NSF Grants OAC-2442555, CNS-2325956, and the VMware University Research Fund. We also thank the Large-scale Sustainable Systems Group members for their support of this work!

\bibliographystyle{IEEEtran}
\bibliography{bibliography/SC2025,bibliography/pearc2025,bibliography/zccloud2024}

\appendix

Table \ref{tab:top500_carbon} shows the operational and embodied carbon footprint of the Top 500 systems. The carbon values based on different data sources for system attributes are shown, such as top500.org; top500.org + other public data sources; top500.org + other public data sources + interpolated values. 

It is interesting to note that systems ranked next to each other can have significant differences in carbon emissions. For example, there is a difference of 4.3x in the operational carbon emissions between LUMI and Leonardo, due to the difference in average carbon intensity and power consumption. Similarly, the embodied carbon emissions of Frontier are 2.6x higher than those of El Capitan, due to differences in hardware configurations such as accelerator models and storage capacity.

\clearpage
\begin{table}[htbp]
\onecolumn
\caption{Operational and embodied carbon footprint of Top 500 systems.}
\twocolumn
\label{tab:top500_carbon}
\begin{filecontents*}{results/top500_final_carbon_p1.csv}
TOP500 Rank,System Name,OC top500.org,OC +public info,OC +interpolated,EM top500.org,EM +public info,EM +interpolated
1,El Capitan,71590,55360,55360,,51561,51561
2,Frontier,59552,60041,60041,,133225,133225
3,Aurora,93656,108479,108479,,138495,138495
4,Eagle,,3049,3049,,,55495
5,HPC6,18386,18386,18386,,14493,14493
6,Supercomputer Fugaku,97058,75642,75642,51058,51058,51058
7,Alps,2047,2047,2047,,15832,15832
8,LUMI,3785,3785,3785,,19935,19935
9,Leonardo,16285,16285,16285,,19359,19359
10,Tuolumne,8196,5727,5727,,5334,5334
11,MareNostrum 5 ACC,3614,3614,3614,,7975,7975
12,Eos NVIDIA DGX SuperPOD,,7667,7667,,2292,2292
13,Venado,4021,4029,4029,,,10043
14,Sierra,18002,11833,11833,,31396,31396
15,Sunway TaihuLight,54944,54944,54944,,,7252
16,CHIE-3,,8370,8370,,,7444
17,CHIE-2,,4475,4475,,,7391
18,JETI - JUPITER Exascale Transition Instrument,2992,2992,2992,,1916,1916
19,Perlmutter,7127,4685,4685,,7051,7051
20,El Dorado,2686,2910,2910,,1778,1778
21,Gefion,2061,1610,1610,,3397,3397
22,CEA-HE,436,436,436,,1337,1337
23,Selene,6404,6404,6404,,35285,35285
24,Tianhe-2A,66064,66064,66064,,7503,7503
25,SuperPOD,,299,299,,,6117
26,Explorer-WUS3,,,8394,,,6551
27,Jean Zay H100,,451,451,,2297,2297
28,Miyabi-G,3192,3722,3722,1020,2134,2134
29,ISEG,2433,703,703,,,5568
30,Adastra,326,326,326,,1882,1882
31,Ubilink,3923,3923,3923,,804,804
32,Reindeer,,203,203,,,4481
33,JUWELS Booster Module,4027,4027,4027,,4753,4753
34,Israel-1,,5259,5259,,1019,1019
35,MareNostrum 5 GPP,4999,4999,4999,20546,22614,22614
36,TSUBAME4.0,2648,3087,3087,,2917,2917
37,GMO GPU Cloud,,1714,1714,,823,823
38,Shaheen III - CPU,20177,20177,20177,16051,12874,12874
39,HPC5,4894,4894,4894,,3163,3163
40,Sejong,,2931,2931,,,7028
41,kakaocloud,,411,411,,,5470
42,Voyager-EUS2,,,5203,,,5725
43,Crossroads,15208,15237,15237,18707,19468,19468
44,PITAGORA,,183,183,,,5068
45,Setonix – GPU,1807,1319,1319,,1091,1091
46,Discovery 5,2333,2333,2333,,,5924
47,Polaris,,2826,2826,,4478,4478
48,SSC-21,,284,284,,,5423
49,rzAdams,939,636,636,,593,593
50,MUSICA Phase 1,,832,832,,2261,2261
51,Capella,1015,1015,1015,,1399,1399
52,Frontera,7946,8896,8896,10951,13693,13693
53,Yep1,,107,107,,,3844
54,BioHive-2,,839,839,,251,251
55,CEA-HF,1755,1755,1755,9828,10061,10061
56,Dammam-7,,1129,1129,,,4569
57,TAIPEI-1,,165,165,,19,19
58,Wisteria/BDEC-01 (Odyssey),4765,5555,5555,2467,2467,2467
59,DeepL Mercury,,12,12,,,3902
60,Chervonenkis,,1926,1926,,2273,2273
61,Piz Daint,685,685,685,,2750,2750
62,ARCHER2,2190,2292,2292,7833,8933,8933
63,Titan,,249,249,,,4313
64,SuperMUC-NG,6981,3550,3550,5703,4907,4907
65,Dawn,,690,690,,2791,2791
66,Hawk,8915,8915,8915,7637,5490,5490
67,Ghawar-1,8496,8496,8496,7805,7805,7805
68,Frontier TDS,747,813,813,,1813,1813
69,Helios GPU,1561,1561,1561,,648,648
70,Pégaso,746,746,746,,,4686
71,Kestrel GPU,,1535,1535,,1834,1834
72,Lassen,,1991,1991,,3390,3390
73,Guru,9452,9452,9452,6740,6740,6740
74,Maru,9452,9452,9452,6740,6740,6740
75,PANGEA III,484,484,484,,,4479
\end{filecontents*}
\pgfplotstabletypeset[
    string type, 
    col sep=comma, 
    columns={TOP500 Rank,System Name,{OC top500.org},{OC +public info},{OC +interpolated},{EM top500.org},{EM +public info},{EM +interpolated}},
    columns/{TOP500 Rank}/.style={column type=|c|},
    columns/{System Name}/.style={column type=P|},
    columns/{OC top500.org}/.style={column type=C|, column name={top500.org}},
    columns/{OC +public info}/.style={column type=C|, column name={+public info}},
    columns/{OC +interpolated}/.style={column type=C|, column name={+interpolated}},
    columns/{EM top500.org}/.style={column type=C|, column name={top500.org}},
    columns/{EM +public info}/.style={column type=C|, column name={+public info}},
    columns/{EM +interpolated}/.style={column type=C|, column name={+interpolated}},
    every head row/.style={
        before row=
        \toprule
        &  & \multicolumn{3}{c|}{Operational Carbon (MT CO2e)} &  \multicolumn{3}{c|}{Embodied Carbon (MT CO2e)}\\,
        after row=
            \toprule
            \cmidrule(lr){1-8}
    },
    every last row/.style={after row=\bottomrule}
    ]{results/top500_final_carbon_p1.csv}
\end{table}
\clearpage
\begin{table}[htbp]
\begin{filecontents*}{results/top500_final_carbon_p2.csv}
TOP500 Rank,System Name,OC top500.org,OC +public info,OC +interpolated,EM top500.org,EM +public info,EM +interpolated
76,AOBA-S,4508,4681,4681,,,4459
77,HiPerGator AI,1410,2189,2189,,9605,9605
78,SuperMUC-NG Phase 2,1657,1657,1657,,1783,1783
79,Helma,,771,771,,3486,3486
80,TOKI-SORA,2618,2618,2618,1850,1850,1850
81,Pioneer-WUS2,,,1625,,,3445
82,Pioneer-WEU,,,1320,,,3343
83,Pioneer-EUS,,,1233,,,2717
84,Pioneer-SCUS,,,1190,,,2811
85,Kempner,,2390,2390,,4679,4679
86,Galushkin,,1316,1316,,1553,1553
87,kakaocloud,,178,178,,,2434
88,Virga,,227,227,,,2343
89,Santos Dumont,225,225,225,,,2306
90,ASPIRE 2A+,,857,857,,711,711
91,Nurion,5168,5121,5121,3678,5606,5606
92,THE CRUST 2.5,,157,157,,,1882
93,Proxima,,1380,1380,,716,716
94,Snellius Phase 3 GPU,545,545,545,,724,724
95,,2937,2937,2937,1480,1480,1480
96,,2937,2937,2937,1480,1480,1480
97,NHN CLOUD GWANGJU AI,,76,76,,,2158
98,Lyapunov,,1344,1344,,1019,1019
99,Lem,,,1577,,,2237
100,HPC4,2869,2869,2869,,,2388
101,Tera-1000-2,1124,1124,1124,3623,3623,3623
102,Kuma,79,79,79,,1495,1495
103,Christofari Neo,,120,120,,,2572
104,Carpenter,2662,3739,3739,5055,4240,4240
105,Kestrel CPU,3470,5154,5154,6494,4920,4920
106,NeXXt AI Factory,,47,47,,,3066
107,DeepL Ceres,75,75,75,,,3010
108,43,592,592,592,,,3162
109,,2478,2478,2478,1233,1233,1233
110,Stampede2,2808,2341,2341,2368,2835,2835
111,Opera Iceland KEF-1 SuperPOD,,4,4,,,3667
112,MeluXina - Accelerator Module,689,689,689,,1733,1733
113,KT DGX SuperPOD,,128,128,,,4016
114,Derecho CPU partition,3350,5946,5946,3838,3863,3863
115,Levante,3515,3413,3413,4270,4146,4146
116,SNL CTS-2 Hops,,109,109,,,3820
117,Cactus,2788,3401,3401,3581,6811,6811
118,Dogwood,2788,3232,3232,3581,6811,6811
119,Earth Simulator -SX-Aurora TSUBASA,4515,4515,4515,,,3744
120,Amazon EC2 Instance Cluster us-east-1a,5573,5573,5573,5895,5895,5895
121,Dardel GPU,45,45,45,,415,415
122,ROMEO-2025,57,57,57,,847,847
123,NVIDIA Cambridge-1 DGX SuperPOD,,31,31,,11,11
124,SNL CTS-2 Stout,3609,3616,3616,4816,4816,4816
125,SNL/NNSA CTS-2 Amber,2561,2566,2566,4794,4794,4794
126,,,114,114,,,2204
127,DGX SuperPOD,,112,112,,,2383
128,,2768,2768,2768,,,2537
129,JURECA Data Centric Module,877,877,877,,975,975
130,Gadi,,910,910,,278,278
131,Goethe-NHR,446,446,446,,893,893
132,Aitken,4865,3198,3198,3212,5445,5445
133,Taiwania 2,2934,2934,2934,,3568,3568
134,Dragão,681,681,681,,,2358
135,Raven-GPU,861,861,861,,1027,1027
136,AIRAWAT - PSAI,,1544,1544,,2955,2955
137,Tuwaiq-1,,,1695,,,2791
138,C24,2485,2485,2485,3595,3595,3595
139,AiMOS,1239,1066,1066,,1250,1250
140,CATT,1784,2506,2506,3595,3595,3595
141,Taranis,592,592,592,3223,3223,3223
142,LANTA,1088,1088,1088,,895,895
143,Roxy,3171,3171,3171,7349,7349,7349
144,HoreKa-Green,819,819,819,,789,789
145,Narwhal,2927,4385,4385,3021,3019,3019
146,42,805,805,805,,526,526
147,Minerva,414,356,356,,557,557
\end{filecontents*}
\pgfplotstabletypeset[
    string type, 
    col sep=comma, 
    columns={TOP500 Rank,System Name,{OC top500.org},{OC +public info},{OC +interpolated},{EM top500.org},{EM +public info},{EM +interpolated}},
    columns/{TOP500 Rank}/.style={column type=|c|},
    columns/{System Name}/.style={column type=P|},
    columns/{OC top500.org}/.style={column type=C|, column name={top500.org}},
    columns/{OC +public info}/.style={column type=C|, column name={+public info}},
    columns/{OC +interpolated}/.style={column type=C|, column name={+interpolated}},
    columns/{EM top500.org}/.style={column type=C|, column name={top500.org}},
    columns/{EM +public info}/.style={column type=C|, column name={+public info}},
    columns/{EM +interpolated}/.style={column type=C|, column name={+interpolated}},
    every head row/.style={
        before row=
        \toprule
        &  & \multicolumn{3}{c|}{Operational Carbon (MT CO2e)} &  \multicolumn{3}{c|}{Embodied Carbon (MT CO2e)}\\,
        after row=
            \toprule
            \cmidrule(lr){1-8}
    },
    every last row/.style={after row=\bottomrule}
    ]{results/top500_final_carbon_p2.csv}
\end{table}
\clearpage
\begin{table}[htbp]
\begin{filecontents*}{results/top500_final_carbon_p3.csv}
TOP500 Rank,System Name,OC top500.org,OC +public info,OC +interpolated,EM top500.org,EM +public info,EM +interpolated
148,Plasma Simulator,5032,5214,5214,,,5469
149,Leonardo-CPU,2430,2430,2430,4870,27808,27808
150,Underhill,1243,1243,1243,3099,3099,3099
151,Robert,1243,1243,1243,3099,3099,3099
152,Belenos,586,586,586,3223,3223,3223
153,alpha ONE,,70,70,,,4881
154,PupMaya,3084,3084,3084,5221,5221,5221
155,Isambard-AI phase 1,102,102,102,,327,327
156,Nscale Svartisen,,20,20,,285,285
157,Gautschi AI,407,571,571,,247,247
158,Artemis,1806,1806,1806,,34,34
159,LLNL/NNSA CTS-2 Bengal,1626,1249,1249,3044,1949,1949
160,LLNL CTS-2 Dane,1870,1719,1719,3500,2682,2682
161,,1402,1402,1402,3888,3888,3888
162,JOLIOT-CURIE ROME,508,508,508,,,2141
163,Gaia,414,414,414,,,2331
164,Theta,2279,2647,2647,1922,5038,5038
165,Karolina GPU partition,1048,1048,1048,,732,732
166,,2149,2149,2149,,,2668
167,Camphor 3,3333,3333,3333,3551,1143,1143
168,HPCF-AD,2608,2608,2608,2686,2851,2851
169,HPCF-AC,2608,2608,2608,2686,2851,2851
170,HPCF-AB,2608,2608,2608,2686,2851,2851
171,HPCF-AA,2608,2608,2608,2686,2851,2851
172,Christofari,,83,83,,,2658
173,Flow,1047,1085,1085,740,732,732
174,MareNostrum,1418,1418,1418,2857,2900,2900
175,SPR-HBM,2570,2570,2570,3396,3396,3396
176,ONYX,3379,4745,4745,3543,3543,3543
177,LUMI-C,647,647,647,2360,3462,3462
178,Genkai Node Group A,3140,2102,2102,3247,3247,3247
179,DAIDC,,83,83,,,2637
180,JUWELS Module 1,3106,3106,3106,2135,2033,2033
181,,,265,265,,,2480
182,Honhai Super AI Computing Center 1,,14835,14835,,2466,2466
183,SQUID - CPU Nodes,3967,3092,3092,2541,2541,2541
184,Lise,2029,1569,1569,3116,2491,2491
185,Kraken-Fregata,,549,549,,289,289
186,Pleiades,10666,7011,7011,8596,2093,2093
187,Emmy+,2002,1556,1556,3075,2281,2281
188,Arka,5507,6374,6374,4691,4691,4691
189,Arunika,5507,4945,4945,3284,2974,2974
190,Narval (GPU),299,76,76,,1961,1961
191,Sawtooth,3466,3071,3071,2843,2843,2843
192,,1289,1289,1289,617,617,617
193,UltraViolet,,27,27,,,2231
194,Grete,469,469,469,,539,539
195,Electra,4080,2682,2682,2353,2832,2832
196,Raven,1817,1739,1739,2661,2547,2547
197,Snellius Phase 2 CPU,1473,1473,1473,2759,3056,3056
198,rzVernal,,138,138,,254,254
199,Clementina XXI,,537,537,,,2645
200,MAHTI,570,570,570,,,2856
201,NA5,2847,2847,2847,3262,3262,3262
202,NA5A,2847,2847,2847,3262,3262,3262
203,Pangea,1468,1468,1468,7574,7574,7574
204,Berzelius,,59,59,,2274,2274
205,DiRAC Tursa,212,212,212,,853,853
206,TX-GAIA (Green AI Accelerator),3475,3475,3475,,,2145
207,Dhabi Supercomputer,373,373,373,,,2033
208,Topaze-gpu,,15,15,,22,22
209,Dardel CPU,149,156,156,1712,2627,2627
210,Athena,,863,863,,475,475
211,Cardinal GPU Partition,200,232,232,,1069,1069
212,,,30,30,,,1663
213,BioHive-1,,105,105,,32,32
214,Flow Type II subsystem,,1078,1078,,3024,3024
215,Joule 3.0,1140,1275,1275,2953,3142,3142
216,Eagle,2033,2684,2684,1589,1589,1589
217,Cheyenne,4180,7371,7371,2329,2613,2613
218,Betzy,76,76,76,1880,1880,1880
219,Derecho GPU partition,409,721,721,,417,417
220,Tioga,,116,116,,214,214
221,Discoverer,1549,1549,1549,1578,1633,1633
222,JEDI,154,154,154,,281,281
223,Jean Zay,,210,210,,1265,1265
\end{filecontents*}
\pgfplotstabletypeset[
    string type, 
    col sep=comma, 
    columns={TOP500 Rank,System Name,{OC top500.org},{OC +public info},{OC +interpolated},{EM top500.org},{EM +public info},{EM +interpolated}},
    columns/{TOP500 Rank}/.style={column type=|c|},
    columns/{System Name}/.style={column type=P|},
    columns/{OC top500.org}/.style={column type=C|, column name={top500.org}},
    columns/{OC +public info}/.style={column type=C|, column name={+public info}},
    columns/{OC +interpolated}/.style={column type=C|, column name={+interpolated}},
    columns/{EM top500.org}/.style={column type=C|, column name={top500.org}},
    columns/{EM +public info}/.style={column type=C|, column name={+public info}},
    columns/{EM +interpolated}/.style={column type=C|, column name={+interpolated}},
    every head row/.style={
        before row=
        \toprule
        &  & \multicolumn{3}{c|}{Operational Carbon (MT CO2e)} &  \multicolumn{3}{c|}{Embodied Carbon (MT CO2e)}\\,
        after row=
            \toprule
            \cmidrule(lr){1-8}
    },
    every last row/.style={after row=\bottomrule}
    ]{results/top500_final_carbon_p3.csv}
\end{table}
\clearpage
\begin{table}[htbp]
\begin{filecontents*}{results/top500_final_carbon_p4.csv}
TOP500 Rank,System Name,OC top500.org,OC +public info,OC +interpolated,EM top500.org,EM +public info,EM +interpolated
224,Wisteria/BDEC-01 (Aquarius),597,696,696,,350,350
225,Unaizah-4,,,558,,,1371
226,Atlas,395,395,395,,,1487
227,ARF ACC,246,246,246,,491,491
228,Pegasus,423,423,423,,1593,1593
229,Advanced Computing System(PreE),1358,1358,1358,,,1862
230,CRONOS,434,434,434,2860,2860,2860
231,SNL/NNSA CTS-1 Manzano,2323,2328,2328,2504,2504,2504
232,NA7A,2201,2201,2201,2522,2522,2522
233,NA7,2201,2201,2201,2522,2522,2522
234,NA2C,2116,2116,2116,2923,2923,2923
235,NA2B,2116,2116,2116,2923,2923,2923
236,NA2,2116,2116,2116,2923,2923,2923
237,Berzelius2,,33,33,,2387,2387
238,Noctua 2,1983,1983,1983,1640,1889,1889
239,NA6B,2113,2113,2113,2918,2918,2918
240,NA6A,2113,2113,2113,2918,2918,2918
241,NA6,2113,2113,2113,2918,2918,2918
242,,,97,97,,,2484
243,Alpha,,20,20,,,2494
244,Raphael,217,217,217,,,2555
245,JOLIOT-CURIE SKL,324,324,324,1482,2265,2265
246,NA12A,1987,1987,1987,2276,2276,2276
247,NA12,1987,1987,1987,2276,2276,2276
248,Abel,4356,4356,4356,2318,2318,2318
249,Warhawk,1376,1388,1388,1420,2677,2677
250,Alex,409,409,409,,767,767
251,Raider,,44,44,,81,81
252,DGX Saturn V,,84,84,,,1993
253,NA9B,2122,2122,2122,2431,2431,2431
254,NA9A,2122,2122,2122,2431,2431,2431
255,NA9,2122,2122,2122,2431,2431,2431
256,NA9Z,2122,2122,2122,2431,2431,2431
257,Deucalion,258,258,258,524,2084,2084
258,,1648,1648,1648,2033,3136,3136
259,,1648,1648,1648,2033,3136,3136
260,NA1,2025,2025,2025,2320,2320,2320
261,Nautilus,,223,223,,279,279
262,Spartan GPU Partitions,,74,74,,89,89
263,Gemini,207,207,207,,,1987
264,VEGA HPC CPU,882,882,882,1343,5486,5486
265,Delta,,527,527,,681,681
266,hotlum,181,181,181,1416,1416,1416
267,CLAIX-2023 (GPU segment),356,356,356,,417,417
268,Pratyush,6030,6980,6980,2438,2911,2911
269,,,112,112,,,1918
270,LLNL Ruby,1220,790,790,2571,2057,2057
271,Orion,1281,1835,1835,2168,2220,2220
272,IARA,,8,8,,,1799
273,,1659,1673,1673,1551,1565,1565
274,Grete Phase 3,182,182,182,,195,195
275,NA10B,1952,1952,1952,2236,2236,2236
276,NA10A,1952,1952,1952,2236,2236,2236
277,NA10,1952,1952,1952,2236,2236,2236
278,Joule 2.0,,224,224,,169,169
279,Snellius Phase 1 GPU,378,378,378,,385,385
280,Niagara,883,586,586,1525,2577,2577
281,Fritz,1534,1534,1534,1658,1658,1658
282,,3628,3628,3628,,,1788
283,,3628,3628,3628,,,1743
284,Anvil,1917,2692,2692,1546,1553,1553
285,NOBZ1,426,426,426,2423,2423,2423
286,AIC1,1643,1643,1643,2423,2423,2423
287,B1A,1429,1429,1429,2423,2423,2423
288,Geely Wise Star-Dubhe,1601,1601,1601,2029,2029,2029
289,Altair,4084,4084,4084,2221,2034,2034
290,Forerunner 1,2262,2262,2262,1769,1769,1769
291,ATOS THX.A.B,30,30,30,,,2084
292,Unizah-II,2443,2443,2443,3599,3599,3599
293,Ohtaka,2181,2181,2181,2350,2350,2350
294,BA1A,2196,2196,2196,2140,2140,2140
295,CARO,1245,1245,1245,1908,1908,1908
296,davinci-1,,63,63,,,1503
297,Tenaya,,87,87,,161,161
298,SEAS H100,195,168,168,,398,398
299,Cedar (GPU),298,152,152,,773,773
300,Molecular Simulator,1512,1512,1512,1292,1292,1292
301,ASPIRE 2A (GPU Partition),,397,397,,326,326
\end{filecontents*}
\pgfplotstabletypeset[
    string type, 
    col sep=comma, 
    columns={TOP500 Rank,System Name,{OC top500.org},{OC +public info},{OC +interpolated},{EM top500.org},{EM +public info},{EM +interpolated}},
    columns/{TOP500 Rank}/.style={column type=|c|},
    columns/{System Name}/.style={column type=P|},
    columns/{OC top500.org}/.style={column type=C|, column name={top500.org}},
    columns/{OC +public info}/.style={column type=C|, column name={+public info}},
    columns/{OC +interpolated}/.style={column type=C|, column name={+interpolated}},
    columns/{EM top500.org}/.style={column type=C|, column name={top500.org}},
    columns/{EM +public info}/.style={column type=C|, column name={+public info}},
    columns/{EM +interpolated}/.style={column type=C|, column name={+interpolated}},
    every head row/.style={
        before row=
        \toprule
        &  & \multicolumn{3}{c|}{Operational Carbon (MT CO2e)} &  \multicolumn{3}{c|}{Embodied Carbon (MT CO2e)}\\,
        after row=
            \toprule
            \cmidrule(lr){1-8}
    },
    every last row/.style={after row=\bottomrule}
    ]{results/top500_final_carbon_p4.csv}
\end{table}
\clearpage
\begin{table}[htbp]
\begin{filecontents*}{results/top500_final_carbon_p5.csv}
TOP500 Rank,System Name,OC top500.org,OC +public info,OC +interpolated,EM top500.org,EM +public info,EM +interpolated
302,NA4,1612,1612,1612,2226,2226,2226
303,NA4A,1612,1612,1612,2226,2226,2226
304,Topaz,12524,17589,17589,1754,2845,2845
305,,846,846,846,,,1736
306,Topaze-cpu,171,171,171,1342,1342,1342
307,Damson,985,985,985,2327,2327,2327
308,LLNL/NNSA CTS-1 MAGMA,2147,1411,1411,1595,1845,1845
309,Mustang,1143,1143,1143,1031,1031,1031
310,NA13A,1653,1653,1653,1894,1894,1894
311,NA13,1653,1653,1653,1894,1894,1894
312,Big Red 200 GPU Partition,,397,397,,326,326
313,HPC2,2666,2666,2666,,,1738
314,Dream-AI,,54,54,,,1727
315,Fênix,281,281,281,,,1797
316,Toubkal,2793,2793,2793,2140,1774,1774
317,,1270,1270,1270,2154,2154,2154
318,B5A,1270,1270,1270,2154,2154,2154
319,AID1,1460,1460,1460,2154,2154,2154
320,NOJP1,1704,1704,1704,2154,2154,2154
321,NONY1,66,66,66,2154,2154,2154
322,NO6,1270,1270,1270,2154,2154,2154
323,Lichtenberg II (Phase 1),1575,1575,1575,1512,1382,1382
324,,15314,15314,15314,6937,6937,6937
325,CLAIX-2023 (CPU segment),1544,1544,1544,1896,3035,3035
326,B2G,1062,1062,1062,3230,3230,3230
327,A1C,1958,1958,1958,3230,3230,3230
328,A5A,1394,1394,1394,3230,3230,3230
329,A1A,1394,1394,1394,3230,3230,3230
330,A1A,1394,1394,1394,3230,3230,3230
331,NO4,1394,1394,1394,3230,3230,3230
332,NA8B,1590,1590,1590,2196,2196,2196
333,NA8A,1590,1590,1590,2196,2196,2196
334,NA8,1590,1590,1590,2196,2196,2196
335,NA8,1590,1590,1590,2196,2196,2196
336,NA8Z,1590,1590,1590,2196,2196,2196
337,Thunder,11664,11664,11664,,,1607
338,HoreKa-Teal,113,113,113,,183,183
339,IronMan,,12,12,,,1471
340,Komondor,188,188,188,,222,222
341,Lichtenberg II (Phase 2),920,920,920,1826,1826,1826
342,VEGA HPC GPU,,213,213,,520,520
343,,2037,2037,2037,2335,2335,2335
344,Alps,183,183,183,1399,1432,1432
345,Helios CPU,2236,2236,2236,1376,1353,1353
346,Medusa,,881,881,,404,404
347,Numerical Materials Simulator,2244,2244,2244,1878,1501,1501
348,Hera,1159,964,964,1961,2106,2106
349,Camphor 2,2428,2428,2428,648,1334,1334
350,Jean Zay,162,76,76,2090,985,985
351,Barnard,1223,1223,1223,1998,1998,1998
352,Palmetto2,,200,200,,233,233
353,Mistral,2547,2547,2547,3398,762,762
354,ARF,1029,1046,1046,1573,1581,1581
355,NO3,1336,1336,1336,3096,3096,3096
356,,1973,1973,1973,2261,2261,2261
357,HHH1,1982,1982,1982,2827,2827,2827
358,III1,1958,1958,1958,2793,2793,2793
359,Tetralith,136,113,113,1525,2016,2016
360,,53,53,53,1399,1399,1399
361,JJJ1,1935,1935,1935,2759,2759,2759
362,,1930,1930,1930,2212,2212,2212
363,A9A,1301,1301,1301,3015,3015,3015
364,A6A,1301,1301,1301,3015,3015,3015
365,Jean,976,1132,1132,1413,1900,1900
366,KKK1,1911,1911,1911,2726,2726,2726
367,,1096,1096,1096,1720,1720,1720
368,LLL1,1887,1887,1887,2692,2692,2692
369,Henri,107,107,107,,,2693
370,,1609,1609,1609,3028,3028,3028
371,A8A,1278,1278,1278,2961,2961,2961
372,A8A,1278,1278,1278,2961,2961,2961
373,A3A,1278,1278,1278,2961,2961,2961
374,A14A,1278,1278,1278,2961,2961,2961
375,A14A,1278,1278,1278,2961,2961,2961
376,,2142,2142,2142,3601,3601,3601
377,MMM1,1864,1864,1864,2658,2658,2658
\end{filecontents*}
\pgfplotstabletypeset[
    string type, 
    col sep=comma, 
    columns={TOP500 Rank,System Name,{OC top500.org},{OC +public info},{OC +interpolated},{EM top500.org},{EM +public info},{EM +interpolated}},
    columns/{TOP500 Rank}/.style={column type=|c|},
    columns/{System Name}/.style={column type=P|},
    columns/{OC top500.org}/.style={column type=C|, column name={top500.org}},
    columns/{OC +public info}/.style={column type=C|, column name={+public info}},
    columns/{OC +interpolated}/.style={column type=C|, column name={+interpolated}},
    columns/{EM top500.org}/.style={column type=C|, column name={top500.org}},
    columns/{EM +public info}/.style={column type=C|, column name={+public info}},
    columns/{EM +interpolated}/.style={column type=C|, column name={+interpolated}},
    every head row/.style={
        before row=
        \toprule
        &  & \multicolumn{3}{c|}{Operational Carbon (MT CO2e)} &  \multicolumn{3}{c|}{Embodied Carbon (MT CO2e)}\\,
        after row=
            \toprule
            \cmidrule(lr){1-8}
    },
    every last row/.style={after row=\bottomrule}
    ]{results/top500_final_carbon_p5.csv}
\end{table}
\clearpage
\begin{table}[htbp]
\begin{filecontents*}{results/top500_final_carbon_p6.csv}
TOP500 Rank,System Name,OC top500.org,OC +public info,OC +interpolated,EM top500.org,EM +public info,EM +interpolated
378,Karolina CPU partition,1695,1695,1695,1007,1007,1007
379,NNN1,1840,1840,1840,2625,2625,2625
380,Freeman,1346,1890,1890,1729,1729,1729
381,,1009,1009,1009,1408,1408,1408
382,SuperMUC Phase 2,3380,3380,3380,1561,1561,1561
383,OOO1,1817,1817,1817,2591,2591,2591
384,,1171,1171,1171,1444,1444,1444
385,,1171,1171,1171,1444,1444,1444
386,JFRS-1,2264,2264,2264,1032,1032,1032
387,,1823,1823,1823,2089,2089,2089
388,Lucia,111,111,111,,236,236
389,,45,45,45,2177,2177,2177
390,,1793,1793,1793,2557,2557,2557
391,Spartan3,143,143,143,1118,1118,1118
392,ZZ2,1802,1802,1802,2594,2594,2594
393,,1780,1780,1780,2270,2270,2270
394,,,328,328,,,2166
395,,1769,1769,1769,2524,2524,2524
396,VSC-4,330,330,330,1080,1565,1565
397,SNL/NNSA CTS-1 Attaway,1815,1819,1819,1807,1807,1807
398,ZZ3,1780,1780,1780,2563,2563,2563
399,,1746,1746,1746,2490,2490,2490
400,Arka AI/ML,,22,22,,,2306
401,ZZ4,1759,1759,1759,2532,2532,2532
402,,1722,1722,1722,2456,2456,2456
403,Banting,383,383,383,1635,1635,1635
404,Hypercluster,,36,36,,,2856
405,A14A,470,470,470,2746,2746,2746
406,NO2,1185,1185,1185,2746,2746,2746
407,ZZ5,1737,1737,1737,2501,2501,2501
408,ZZ6,1699,1699,1699,2423,2423,2423
409,,3575,3575,3575,6730,6730,6730
410,Grace,630,622,622,1460,4239,4239
411,LLNL/NNSA CTS-1 Jade,32962,21667,21667,1534,1158,1158
412,LLNL CTS-1 Quartz,32962,21667,21667,1534,2684,2684
413,NEA1,4289,4289,4289,10095,10095,10095
414,,1716,1716,1716,1966,1966,1966
415,,1319,1319,1319,1514,1514,1514
416,A7A,1162,1162,1162,2692,2692,2692
417,A7A,1162,1162,1162,2692,2692,2692
418,A1B,1162,1162,1162,2692,2692,2692
419,NO1,1162,1162,1162,2692,2692,2692
420,,1675,1675,1675,2389,2389,2389
421,Cedar (CPU),761,389,389,2369,4314,4314
422,Daley,383,383,383,1635,1635,1635
423,AlphaCentauri,374,374,374,,608,608
424,HPC3,,473,473,,,2259
425,,1673,1673,1673,2625,2625,2625
426,Conesus,990,990,990,1218,1218,1218
427,,1581,1581,1581,2255,2255,2255
428,,1430,1430,1430,2692,2692,2692
429,ASPIRE 2A (CPU Partition),1175,1175,1175,1192,2160,2160
430,,1651,1651,1651,2355,2355,2355
431,Mihir,4254,3820,3820,1709,2039,2039
432,Spartan2,37,37,37,,,1509
433,,1422,1422,1422,1573,301,301
434,,1422,1422,1422,1573,301,301
435,PAI-BSystem,2538,2538,2538,,,964
436,Sol,420,512,512,,1342,1342
437,cascade,3349,3349,3349,,1506,1506
438,Champollion,21,21,21,,133,133
439,Rattler,,52,52,,14,14
440,Adastra 2,13,13,13,,135,135
441,HEMUS,,179,179,,759,759
442,Improv,898,1015,1015,1281,1250,1250
443,occigen2,506,506,506,1773,1947,1947
444,Crash1,86,86,86,1056,1056,1056
445,,1510,1510,1510,2154,2154,2154
446,Excalibur,3547,4109,4109,1589,1589,1589
447,CLAIX (2018),1580,1580,1580,1141,1504,1504
448,Expanse,638,521,521,820,2377,2377
449,,,80,80,,,1496
450,,,80,80,,,1540
451,Lomonosov 2,,1559,1559,,1013,1013
452,Kay,827,944,944,1197,1350,1350
453,B1,1609,1609,1609,1843,1843,1843
454,Centennial,2021,2021,2021,1360,1732,1732
455,Greene-H100,,498,498,,344,344
\end{filecontents*}
\pgfplotstabletypeset[
    string type, 
    col sep=comma, 
    columns={TOP500 Rank,System Name,{OC top500.org},{OC +public info},{OC +interpolated},{EM top500.org},{EM +public info},{EM +interpolated}},
    columns/{TOP500 Rank}/.style={column type=|c|},
    columns/{System Name}/.style={column type=P|},
    columns/{OC top500.org}/.style={column type=C|, column name={top500.org}},
    columns/{OC +public info}/.style={column type=C|, column name={+public info}},
    columns/{OC +interpolated}/.style={column type=C|, column name={+interpolated}},
    columns/{EM top500.org}/.style={column type=C|, column name={top500.org}},
    columns/{EM +public info}/.style={column type=C|, column name={+public info}},
    columns/{EM +interpolated}/.style={column type=C|, column name={+interpolated}},
    every head row/.style={
        before row=
        \toprule
        &  & \multicolumn{3}{c|}{Operational Carbon (MT CO2e)} &  \multicolumn{3}{c|}{Embodied Carbon (MT CO2e)}\\,
        after row=
            \toprule
            \cmidrule(lr){1-8}
    },
    every last row/.style={after row=\bottomrule}
    ]{results/top500_final_carbon_p6.csv}
\end{table}
\clearpage
\begin{table}[htbp]

\begin{filecontents*}{results/top500_final_carbon_p7.csv}
TOP500 Rank,System Name,OC top500.org,OC +public info,OC +interpolated,EM top500.org,EM +public info,EM +interpolated
456,PAI-ASystem,2645,2645,2645,1138,1138,1138
457,C1,1587,1587,1587,1819,1819,1819
458,Big Red 200,719,979,979,923,895,895
459,,3217,3217,3217,6057,6057,6057
460,D1,1576,1576,1576,1807,1807,1807
461,,3172,3172,3172,5973,5973,5973
462,E1,1555,1555,1555,1782,1782,1782
463,Cannon2-A100,211,182,182,,306,306
464,,1439,1439,1439,2053,2053,2053
465,,3128,3128,3128,5889,5889,5889
466,,699,699,699,1050,1050,1050
467,B2F,725,725,725,1615,1615,1615
468,B2E,725,725,725,1615,1615,1615
469,B2B,725,725,725,1615,1615,1615
470,B2D,953,953,953,1615,1615,1615
471,B2C,953,953,953,1615,1615,1615
472,B2A,953,953,953,1615,1615,1615
473,NOG1,898,898,898,1615,1615,1615
474,NON1,725,725,725,1615,1615,1615
475,NOUK1,342,342,342,1615,1615,1615
476,NVIDIA DGX SuperPOD,218,218,218,,,3259
477,ZZ1,5834,5834,5834,14822,14822,14822
478,A12A,1202,1202,1202,2423,2423,2423
479,A11A,1202,1202,1202,2423,2423,2423
480,A17A,54,54,54,2423,2423,2423
481,A17A,124,124,124,2423,2423,2423
482,UK2,375,375,375,2423,2423,2423
483,UK1,375,375,375,2423,2423,2423
484,B1B,1046,1046,1046,2423,2423,2423
485,A10A,1046,1046,1046,2423,2423,2423
486,A4A,1046,1046,1046,2423,2423,2423
487,Ares,2399,2399,2399,1326,1320,1320
488,,,83,83,,,2207
489,,,83,83,,,2185
490,Baskerville,,104,104,,381,381
491,HoreKa-Blue,1129,1129,1129,1006,1974,1974
492,,,71,71,,,2340
493,,,71,71,,,2442
494,,,71,71,,,2466
495,Makman-3,2314,2314,2314,1260,1260,1260
496,F1,1501,1501,1501,1721,1721,1721
497,,,,1015,,,2823
498,,3038,3038,3038,5720,5720,5720
499,VSC-5,448,448,448,1155,5684,5684
500,Marlyn,502,605,605,1011,949,949
\end{filecontents*}
\pgfplotstabletypeset[
    string type, 
    col sep=comma, 
    columns={TOP500 Rank,System Name,{OC top500.org},{OC +public info},{OC +interpolated},{EM top500.org},{EM +public info},{EM +interpolated}},
    columns/{TOP500 Rank}/.style={column type=|c|},
    columns/{System Name}/.style={column type=P|},
    columns/{OC top500.org}/.style={column type=C|, column name={top500.org}},
    columns/{OC +public info}/.style={column type=C|, column name={+public info}},
    columns/{OC +interpolated}/.style={column type=C|, column name={+interpolated}},
    columns/{EM top500.org}/.style={column type=C|, column name={top500.org}},
    columns/{EM +public info}/.style={column type=C|, column name={+public info}},
    columns/{EM +interpolated}/.style={column type=C|, column name={+interpolated}},
    every head row/.style={
        before row=
        \toprule
        &  & \multicolumn{3}{c|}{Operational Carbon (MT CO2e)} &  \multicolumn{3}{c|}{Embodied Carbon (MT CO2e)}\\,
        after row=
            \toprule
            \cmidrule(lr){1-8}
    },
    every last row/.style={after row=\bottomrule}
    ]{results/top500_final_carbon_p7.csv}
\end{table}
\end{document}